\documentclass[]{aa}  
\usepackage{color }
\usepackage{natbib}
\usepackage{amsmath}
\usepackage{amssymb}
\usepackage{graphicx}
\usepackage{txfonts}
\begin{document}

%\title{Gravitoviscous protoplanetary disks with a dust component.}
\title{The origin of tail-like structures around protoplanetary disks}

\author{Eduard I. Vorobyov\inst{1,2,3}, Alexandr  M. Skliarevskii\inst{2}, Vardan G. Elbakyan\inst{2,4},  Michihiro Takami\inst{5}, Hauyu Baobab Liu\inst{5}, Sheng-Yuan Liu\inst{5} and Eiji Akiyama\inst{6}}
% List of institutions
\institute{ 
University of Vienna, Department of Astrophysics, Vienna, 1180, Austria \\
\email{eduard.vorobiev@univie.ac.at} 
\and
Research Institute of Physics, Southern Federal University, Roston-on-Don, 344090 Russia
\and
Ural Federal University, 51 Lenin Str., 620051 Ekaterinburg, Russia
\and
Lund Observatory, Department of Astronomy and Theoretical Physics, Lund University, Box 43, 22100 Lund, Sweden
\and
Institute of Astronomy and Astrophysics, Academia Sinica, 11F of Astronomy-Mathematics Building, AS/NTU No.1, Sec. 4, Roosevelt
Rd, Taipei 10617, Taiwan, R.O.C.
\and
Institute for the Advancement of Higher Education, Hokkaido University, Kita17, Nishi8, Kita-ku, Sapporo, 060-0817, Japan
}

%   \date{...}

% \abstract{}{}{}{}{} 
% 5 {} token are mandatory
 
  \abstract
   {}
   {We study the origin of tail-like structures recently detected around the disk of SU Aurigae and several FU~Orionis type stars.}
   {Dynamic protostellar disks featuring ejections of gaseous clumps and quiescent protoplanetary disks experiencing a close encounter with an intruder star were modelled using the numerical hydrodynamics code FEOSAD. Both the gas and dust dynamics were taken into account, including dust growth and mutual friction between the gas and dust components. Only plane-of-the-disk encounters were considered.}
   {Ejected clumps produce a unique type of tails that are characterized by a bow-shock shape. These tails owe its origin to the supersonic motion of the ejected clumps through the dense envelope, which often surrounds young gravitationally unstable protostellar disks. The ejected clumps either sit at the head of the tail-like structure or disperse if their mass is insufficient to withstand the head wind of the envelope.   
   On the other hand, close encounters with quiescent protoplanetary disks produce three types of the tail-like structures, which we defined as pre-collisional, post-collisional, and spiral tails. These tails can in principle be distinguished by peculiar features of the gas and dust flow in and around them. We found that the brown-dwarf-mass intruders do not capture circumintruder disks during the encounter, while the sub-solar-mass intruders can acquire appreciable circumintruder disks with elevated dust-to-gas ratios, which can ease their observational detection. However, this is true only for prograde collisions; the retrograde intruders fail to collect an appreciable gas and dust from the disk of the target. The masses of gas in the tails vary in the 0.85--11.8~$M_{\rm Jup}$ limits, while the total mass of dust lies in the 1.75--30.1~$M_\oplus$ range, with the spiral tails featuring the highest masses.  The predicted mass of dust in the model tail-like structures is therefore higher than what was inferred for similar structures in SU~Aur, FU~Ori, and Z~CMa, making their observational detection feasible.  }
   {Tail-like structures around protostellar and protoplanetary disks can act as a smoking gun to infer interesting phenomena, such as clump ejection or close encounters.  particular, the bow-shock morphology of the tails could point to clump ejections as a possible formation mechanism. Further numerical and observational studies are needed to better understand the detectability and properties of the tails. }


   \maketitle

\section{Introduction}
Protoplanetary disks are an important ingredient of the star and planet formation process. They form during the gravitational contraction of
prestellar cloud cores and disperse after several million years of evolution, often leaving behind a planetary system. While directly observing planet formation is difficult, several indirect signatures of this process have recently been identified. 
For instance, observations from the near-infrared to millimeter bands in dust continuum and in scattered light have revealed a variety of disk substructures, such as spiral arcs and arms, clumps, rings, and gaps \citep[e.g.,][]{2013Marel,2015ALMA,2016Perez,2016LiuTakami}, indicating that protoplanetary disks often do not have simple monotonically declining surface density profiles. 
Theoretical interpretations of these substructures imply a presence of already-formed massive planets \citep[e.g.,][]{2012KleyNelson} or/and physical phenomena at work, such as gravitational instability or vortices,  which can assist planet formation \citep[e.g.,][]{2016DongVorobyov, 2017RegalyJuhasz}. 

In addition to a seemingly complicated morphological structure of protoplanetary disks, the forming protostars also show a surprisingly tumultuous behavior. Energetic luminosity outbursts known as FU Orionis-type eruptions \citep[see a review by][]{2014AudardAbraham} imply time-varying mass transport rates through the disk and show that the early evolution of protostars cannot be described by a steady mass accumulation as was originally thought \citep{1977Shu}. Interestingly, some disk substructures may be inherently linked to these energetic luminosity outbursts. For instance, gravitational instability of massive protostellar disks, manifesting itself through spectacular spiral arms and clumps, can trigger luminosity outbursts directly via infall of clumps or indirectly via prompting the magnetorotational instability in the innermost parts of the disk   \citep{2005VorobyovBasu, 2014BaeHartmann, 2017MeyerVorobyov}. Moreover, \citet{2019Kadam} demonstrated that gaseous rings forming in the dead zones of the inner disk can drift slowly towards the star until the rising temperature in their interiors triggers the magnetorotational instability and an associated mass accretion burst.

Although the current efforts in the observations of young protoplanetary disks are concentrated on achieving an ever-sharper view of the disk substructures, lower-resolution observations with a much wider field of view can bring about interesting and unexpected results. For instance, 
\citet{2006Cabrit} detected a tail-like structure trailing from the disk of RW Aurigae~A, which they interpreted as a tidally stripped tail of disk material formed by its companion, RW Aurigae B, during a close fly-by. They also speculated that a warm CO gas in the outer disk regions and elevated mass accretion rates were also a result of close approach. Indeed, close fly-bys of companion, cloudlets, or intruder stars are known to produce tidal tails and  mass accretion bursts \citep[e.g.][]{1992Bonnell, 2010Thies, 2011ForganRice, 2015Dai, 2017VorobyovSteinrueck,2019Dullemond}. The ALMA observations by \citet{2018Rodriguez} further elaborated the encounter scenario for the formation of tail-like structure by suggesting multiple fly-bys in the RW~Aurigae system.  

%{(\bf To Taiwanese colleagues: Do you know more references on tail-like structures?)}.

Recently, \citet{2015deLeon} and  \citet{2019AkiyamaVorobyov} presented the Subaru-HiCIAO and ALMA observations of SU~Auriage showing that this system also has a tail-like structure extending from its disk for a thousand au. Several formation mechanisms, such as close encounters, ejections of gaseous clumps, and gaseous streams from the molecular cloud remnant, were proposed based on the morphology and kinematics of carbon monoxide in the tail. Interestingly, tail-like structures of similar extent have also been identified in several FU Orionis-type objects \citep[e.g., Z CMa, FU Orionis;][]{2016LiuTakami,2018Takami}. However, the exact causal link between the tail-like structures and accretion bursts remains to be understood and is the subject of our future investigations.

In this work we explore the origin of the tail-like structures by means of numerical hydrodynamics simulations.
In particular, we focus on two scenarios: fly-bys of an intruder (sub-)stellar object near an evolved protoplanetary disk and ejections of gaseous clumps from a young and gravitationally unstable protostellar disk.  
We used the Formation and Evolution of Stars and Disks (FEOSAD) code described in detail in \citet{2018VorobyovAkimkin}, modified to include a close encounter with an intruder star according to the method described in \citet{2017VorobyovSteinrueck}. Unlike many previous studies of clump ejection and close encounters \citep[e.g.][]{2003Pfalzner,2007LodatoMeru,2009StamatellosWhitworth,2010ForganRice,2019Cuello}, we also explore the dynamics and growth of the dust component in the tail-like structure, thus making valuable predictions for future observations of dust emission in continuum and scattered light. 
%In addition, we search for possible causal links between tail-like structures and mass accretion bursts typical of FU Orionis-type objects.
The paper is organized as follows. In Sect.~\ref{FEOSAD} we provide a brief overview of the adopted numerical model. Sections~\ref{ejection} and \ref{flybys} describe the tail-like structures formed through clump ejection and close fly-bys, respectively. The main results are summarized in Sect.~\ref{conclude}.

%Your Papers on ejection were focused on the properties of ejected clumps
%May be smoking guns of recent events ...

\section{Numerical model of a gaseous and dusty disk} 
\label{FEOSAD}
To study the origin of tail-like structures we employed the Formation and Evolution Of a Star And its circumstellar Disk (FEOSAD) code described in detail in \citet{2018VorobyovAkimkin}. The FEOSAD code was further modified in \citet{2019VorobyovSkliarevskii} to include the effect of back reaction of dust on gas. We also introduced the central smart cell (CSC) -- a simple model for the innermost disk region which is difficult to resolve in numerical hydrodynamics simulations.  
A close encounter with an intruder star is modelled using the non-inertial frame of reference centered on the target star. The pertinent equations describing the gravitational interaction of the intruder with the target star and its disk were presented in \citet{2017VorobyovSteinrueck}.
We refer the reader to the aforementioned papers for an in-depth description of the code and provide below only a brief overview.

The starting point of each simulation is the gravitational collapse of 
a pre-stellar core. The core has the form of a flattened pseudo-disk, a spatial 
configuration that can be expected in the presence 
of rotation and large-scale magnetic fields \citep{1997Basu}. 
As the collapse proceeds, the inner regions of the core 
spin up and a centrifugally balanced circumstellar disk forms when the inner infalling layers of the core hit the centrifugal barrier near the inner computational boundary.
Depending on the task considered, the radius of the inner boundary can vary from 
1.0~au (collision with an intruder) or 10~au (ejection of a clump from a gravitationally unstable disk). The outer boundary is usually set equal to several thousand au. 
The material that has passed through the inner boundary before the instance of circumstellar disk formation constitutes a seed for the central star, which grows further through accretion from the circumstellar disk.

We use a logarithmically spaced grid in the radial direction and an equally spaced grid in the azimuthal direction. The numerical resolution depends on the problem. When studying collisions with an intruder, the resolution was set to $256\times 256$ grid zones. When studying the ejection of gaseous clumps, the resolution was $1024 \times 1024$ grid zones. A higher resolution in the latter case is needed to resolve the Jeans length in gravitationally fragmenting disks. A higher resolution also allows us to resolve the internal structure of the clumps and avoid their premature tidal destruction \citep{2018VorobyovElbakyan}.

The main physical processes considered in the FEOSAD code when modeling the disk formation and evolution include viscous and shock heating, irradiation by the forming star,  background irradiation,
radiative cooling from the disk surface, friction between the gas and dust components, and self-gravity of gaseous and dusty disks.  The code is written in the thin-disk limit ($r, \phi$), complemented by a calculation of the gas vertical  scale height using an assumption of local hydrostatic equilibrium as described in \citet{VorobyovBasu2009}. The resulting  model has a flared structure (because the disk vertical scale height increases with radius), which guarantees that both the disk and envelope receive a fraction of the irradiation energy  from the central protostar. The mass of the protostar is calculated using  the mass accretion rate through the inner computational boundary. The stellar radius and photospheric luminosity are calculated using the pre-computed stellar evolution tracks obtained with the STELLAR code \citep{2008YorkeBodenheimer}.  The accretion luminosity is calculated using the model's known stellar radius, stellar mass, and mass accretion rate.   

\begin{figure}
\begin{centering}
\includegraphics[width=0.8\columnwidth]{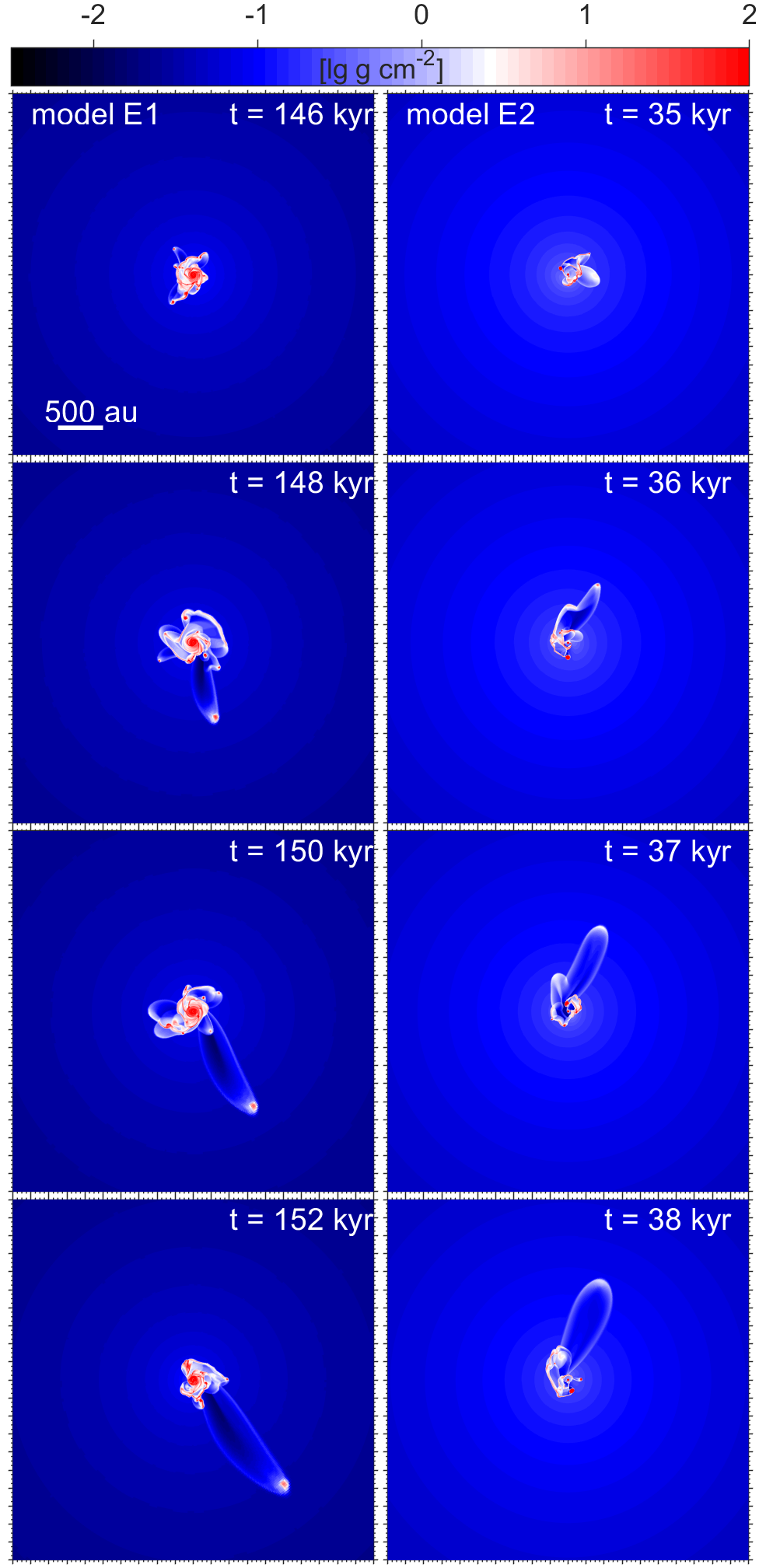}
\par\end{centering}
\caption{\label{fig:1} Gas surface density maps in the inner $4000\times4000~\rm~au^2$ box in model E1 (left column) and E2 (right column) at four different time instances illustrating ejection of clumps from the disk. The time is counted from the formation of the central star.}
\end{figure}

%with a uniform temperature of $T_\mathrm{bg}=20$\,K 
%set equal to the initial temperature of the natal cloud core

%%%%%%%%%%%%% ALL MODELS %%%%%%%%%%%%%%%%

\begin{figure}
\begin{centering}
\includegraphics[width=1\columnwidth]{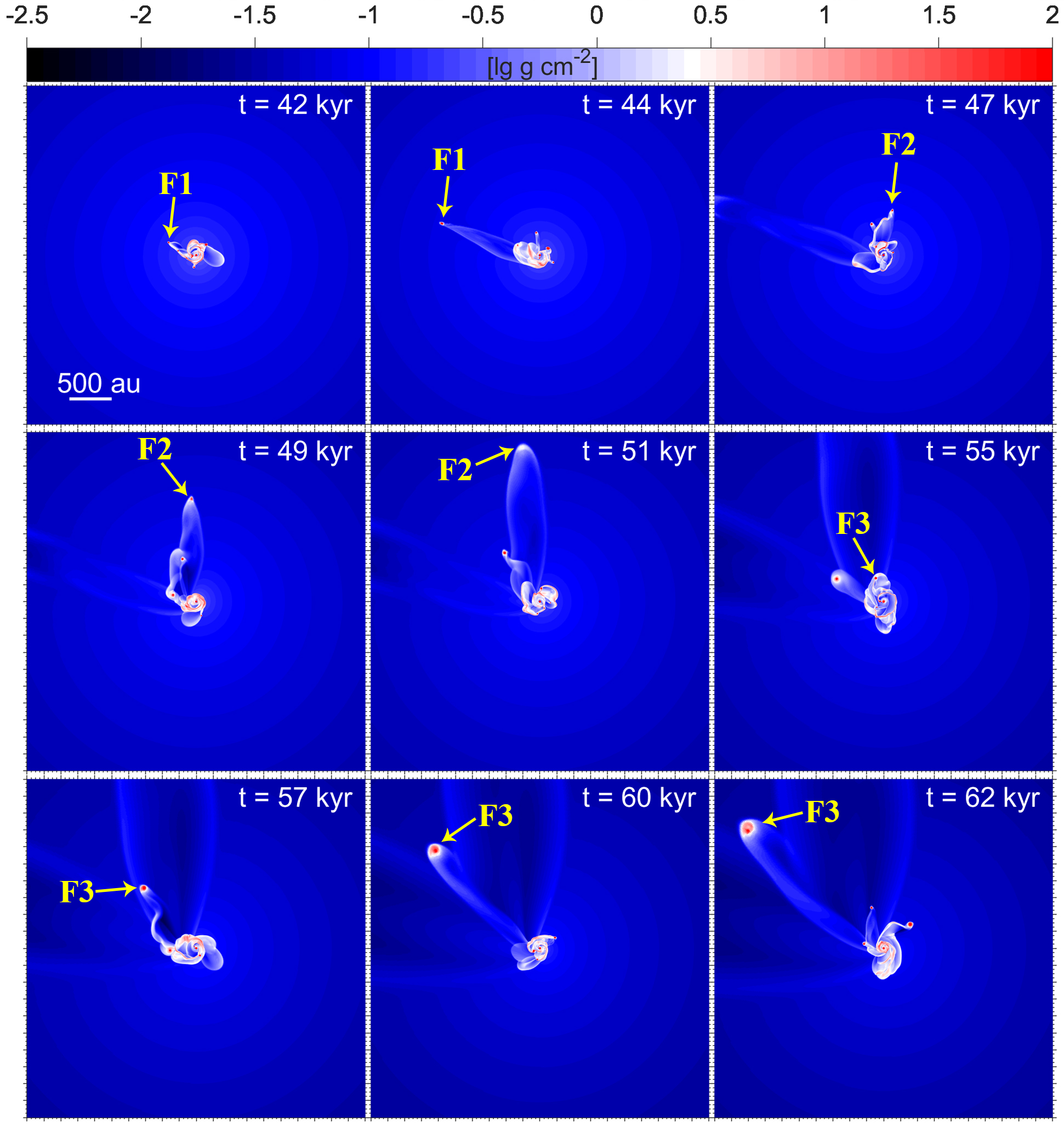}
\par\end{centering}
\protect\caption{\label{fig:2} Gas surface density maps in the inner $4000\times4000~\rm~au^2$ box in model E3 at nine evolutionary times. The time is counted from the formation of the central star. The yellow arrows indicate the ejected clumps, the parameters of which are listed in Table~\ref{tab:2}.}
\end{figure}

In our model, dust consists of two components: small dust ($5\times 10^{-3}$~$\mu$m $<a<$~1.0~$\mu$m) and grown dust ($ 1.0~\mu$m~$\le a<a_{\rm max}$), where $a_{\rm max}$ is a dynamically varying maximum radius of dust grains which depends on the efficiency of radial dust drift and dust growth rate. Small dust is assumed to be coupled to gas, meaning that we only solve the continuity equation for small dust grains, while the dynamics of grown dust is controlled by friction with the gas component and by the total gravitational potential of the star, gaseous and dusty components. Small dust can turn into grown dust, and this process is considered by calculating the dust growth rate and the maximum radius of grown dust as follows. 
The small-to-grown dust conversion rate is defined as
\begin{equation}
\label{GrowthRate}
S(a_{\rm max}) =  {1 \over \Delta t } \Sigma_{\rm d,tot}^n  
{ \int \limits_{a_{\rm max}^n} \limits^{a_{\rm max}^{n+1}} a^{3-p} da \int \limits_{a_{\rm min}} 
\limits^{a_\ast} a^{3-p} da \over \int \limits_{a_{\rm min}} \limits^{a_{\rm max}^n} a^{3-p} da 
\int \limits_{a_{\rm min}} \limits^{a_{\rm max}^{n+1}} a^{3-p} da   },
\end{equation}
where  $\Sigma_{\rm d,tot}$ is the total (small plus grown) surface density of dust, $a_{\rm max}^n$ and $a_{\rm
max}^{n+1}$ the maximum dust radii at the current and next time steps, $a_{\rm min}=0.005~\mu m$ the minimum radius of small dust grains,
$a_\ast=1.0$~$\mu$m a threshold value between small and grown
dust components, and $p=3.5$ describes the distribution function of dust particles per their size (i.e., $N(a)\sim a^{-p}$). The conversion rate $S(a_{\rm max})$ enters the continuity and momentum equations for small dust, grown dust and gas, and accounts for mass and momentum transfer between the gaseous and dusty subsystems as the disk evolves and dust grows \citep[see for detail][]{2018VorobyovAkimkin}. We note that the conversion rate $S(a_{\rm max})$ accounts not only for dust growth, but also for dust fragmentation if $a_{\rm max}^{n+1}$ becomes smaller than $a_{\rm max}^{n}$ in a given computational cell. The latter can occur if the fragmentation barrier (which limits the dust growth) decreases due to a change of physical conditions in the disk.

The maximum size of dust grains in a given cell on the Eulerian grid can change not only due to dust growth or fragmentation through mutual collisions, but also because of dust flow through the cell. The evolution of the maximum radius $a_{\rm max}$ is therefore described as
\begin{equation}
{\partial a_{\rm max} \over \partial t} + (u_{\rm p} \cdot \nabla_p ) a_{\rm max} = \cal{D},
\label{dustA}
\end{equation}
where the growth rate $\cal{D}$ accounts for the change in $a_{\rm max}$ due to coagulation and fragmentation and the second term on the left-hand side account for the change of $a_{\rm max}$ due to dust flow through the cell. $u_{\rm p}$ is the velocity of grown dust in the plane of the disk.  To summarize, $S(a_{\rm max}$) accounts for the conversion of small dust to grown dust and $\cal{D}$ describes the dust growth. We note that the bounds of the small dust component remain fixed. More details on the dust growth scheme, and in particular on the form of $\cal{D}$,  can be found in \citet{2018VorobyovAkimkin}.

As dust grows, the span in the dust sizes ($a_\ast:a_{\rm max}$) and in the corresponding Stokes numbers covered by the grown component may become substantial. However, we only track the dynamics of dust particles with the maximum size by calculating the stopping time for $a_{\rm max}$.  For the chosen power law of $p=3.5$, most of the dust mass is located at the upper end of the dust size distribution, meaning that the dust particles of maximum size are representative for the entire dust mass reservoir. A more rigorous approach  to studying dust dynamics requires the use of multi-bins with narrower ranges of dust sizes and this model is currently under development.

In the next sections we show the spatial distributions of grown dust for some cases, as well as those of gas, which also corresponds to the distributions of small dust.
{ Grown dust can be observed at millimeter wavelengths in particular if these particles grow up to millimeter sizes \citep[e.g.,][]{2016DongVorobyov}, though the corresponding millimeter emission would be a function of temperature as well as the dust mass. It will be our future work to explore the observability of grown dust in our models}. Small dust grains can be observed in the scattered light at optical and near-IR wavelengths, however, they { are more difficult to observe at millimeter wavelengths using ALMA due to smaller opacities \citep{2002Wood,2018Birnstiel} and longer on-source integration times required to achieve signal-to-noise ratios similar to those of grown dust.}  For the tails being actually observed at optical and near-IR wavelengths, these have to be offset from the midplane of the optically thick disk to receive illumination from the star. Although we consider clump ejections, intruders and resultant tails only in the plane of the disk (for which the tail cannot be easily illuminated by the central star  unless the vertical extent of the tail is larger than that of the circumstellar disk), we believe that the simulated distribution of gas (plus small dust) would be useful for discussing how we could discriminate the tails formed through different physical processes using observed morphologies of the tails seen in the optical and near-IR scattered light.

\section{Tail-like structures produced by ejected clumps}
\label{ejection}
The first mechanism for the formation of tail-like structures that we consider is the ejection of gaseous clumps from a massive, gravitationally unstable disk. 
{ The ejection phenomenon was first described in \citet{2007Stamatellos}, who considered an isolated massive disk and found that after its fragmentation the majority of brown-dwarf-mass fragments were releases into the field by interactions among themselves. We note that in Stamatellos et al. the fragments were replaced with sink particles before they were ejected from the disk. The ejection of gaseous clumps (without introducing sink particles)} was studied in detail in \citet{2010VorobyovBasu} and \citet{2016Vorobyov} in the context of the properties of ejected clumps. { It was found that for the ejection of gaseous clumps to occur two conditions have to be fulfilled: 1) the clumps must be well resolved on the numerical grid to avoid artificial dispersal during close encounters and 2) the initial mass and angular momentum of the parental pre-stellar core must be sufficiently high to form massive and extended disks \citep[see fig.~1 in][]{2013Vorobyov}.} In this work, we focus also on the characteristics of the tail-like structures formed in the process of ejection. We also study the dust content in the ejected clumps and tail-like structures thanks to the dust dynamics and growth module in the FEOSAD code.

\begin{table*}
\center
\caption{\label{tab:1}Model parameters}
\begin{tabular}{ccccccccc}
 &  &  &  &  &  &  &  &     \tabularnewline
\hline 
\hline 
Model & $M_{\mathrm{core}}$ & $\beta$ & $T_{\mathrm{init}}$ & $\Omega_{0}$ & $r_{0}$ & $\Sigma_{g,0}$ & $r_{\mathrm{out}}$ & $A$ \tabularnewline
 & [$M_{\odot}$] & [\%] & [K] & [$\mathrm{km\,s^{-1}\,pc^{-1}}$] & [au] & [$\mathrm{g\,cm^{-2}}$] & [pc] \tabularnewline
\hline 
E1 & 1.7 & 0.41 & 15 & 1.12 & 2572 & 0.07 & 0.075 & 1.2 \tabularnewline
%E2 & 1.3 & 0.88 & 15 & 28.45? & 1543 & 0.15 & 0.045 & 1.2 \tabularnewline
E2 & 1.2 & 0.77 & 15 & 3.50 & 1371 & 0.17 & 0.040 & 1.2 \tabularnewline
E3 & 1.0 & 0.77 & 15 & 2.58 & 1474 & 0.13 & 0.045 & 1.2 \tabularnewline
\hline 
\end{tabular}
\center{ \textbf{Notes.} $M_{\mathrm{core}}$ is the initial core
mass, $\beta$ is the ratio of rotational to gravitational energy of the core, $T_{\mathrm{init}}$ is the
initial gas temperature, $\Omega_{0}$ and $\Sigma_{\rm g,0}$ are the angular velocity
and gas surface density at the center of the core, $r_{0}$ is the radius
of the central plateau in the initial core, and $r_{\mathrm{out}}$ is the initial radius of the
core.}
\end{table*}

We consider three models, the initial parameters of which are listed in Table~\ref{tab:1}. These models were chosen so that massive protostellar disks are produced during the collapse of prestellar cores, which are capable of triggering clump ejections. Figure~1 in \citet{2013Vorobyov} indicates that the initial masses and ratios of rotational-to-gravitational energy of the pre-stellar cores have to be greater than $0.6~M_\odot$ and 0.3\%, respectively.  The initial  surface density and angular momentum distributions of the prestellar cores have the following form: 
\begin{equation}
\Sigma_{\rm g}=\frac{r_{0}\Sigma_{\rm g,0}}{\sqrt{r^{2}+r_{0}^{2}}},
\label{eq:sigma}
\end{equation}
\begin{equation}
\Omega=2\Omega_{0}\left(\frac{r_{0}}{r}\right)^{2}\left[\sqrt{1+\left(\frac{r}{r_{0}}\right)^{2}}-1\right].
\label{eq:omega}
\end{equation}
Here, $\Sigma_{\rm g,0}$ and $\Omega_{0}$ are the
gas surface density and angular velocity at the center of the core, $r_{0}=\sqrt{A}c_{\mathrm{s}}^{2}/\pi G\Sigma_{\rm g,0}$
is the radius of the central plateau, where $c_{\mathrm{s}}$ is the initial isothermal sound speed in the core. The initial temperature in the core was set equal to 15~K. This is also the temperature of the external background irradiation. The dust-to-gas ratio was set to the canonical value of 1:100 and all dust was initial in the form of small sub-micron dust grains.

The protostellar disks in our models become gravitationally unstable and fragment in the early stages of their evolution thanks to continuing mass-loading from infalling parental cores.  The fate of the clumps formed through disk fragmentation depends on complicated and often stochastic processes in a strongly unstable disk. The gravitational interaction of the clumps  with each other and with the spiral arms can drive the clumps inward on the star, causing strong mass accretion bursts \citep{2005VorobyovBasu}, or eject them into the interstellar medium via three-body gravitational interaction (two clumps plus the central star), producing free-floating objects \citep{2010VorobyovBasu}. Two examples of clump ejection are shown in Figure~\ref{fig:1} for models E1 and E2.
%Several clump ejection events, as a result of gravitational interaction with each other, are observed in model E1 and E2 during the early evolution of the disk. . 
The time in the top-right corner of each panel shows the age of the system counted from the instance of the central star formation. The disks in both models are young and massive (0.1--0.2~$M_\odot$), hosting a number of already formed clumps.
It is not easy to measure the size of such a highly fragmented disk, but a visual estimate suggests that its radius usually does not exceed several hundreds of au (including clumps on closed orbits).  
The ejected clumps reach a radial distance of about 1000~au within a few kyrs since their ejection from the disk, leaving behind a tail-like structure. The velocity of the ejected clumps in models E1  (at $t=148$~kyr) and in model~E2 (at $t=36$~kyr) are $1~\textrm{km~s}^{-1}$ and $1.6~\textrm{km~s}^{-1}$, respectively. These values are greater than the escape velocity in these models, $\approx$~0.8-0.9~km~s$^{-1}$, meaning that we witness true ejections rather than scattering to a wider orbit.

The ejections occur mostly in the early unstable stages of disk evolution, when the disk is still surrounded by the parental  core. As the clumps transverse through the infalling core, they experience a strong head wind from the surrounding environment. If we consider the bottom row in Figure~\ref{fig:1}, the surrounding gas surface density is $\approx 0.04~\textrm{g~cm}^{-2}$ for the clump in model E1 and $\approx 0.11~\textrm{g~cm}^{-2}$ for the clump in model E2. The sound speed is $c_\textrm{s}=0.4~\textrm{km~s}^{-1}$ for model E1 and $c_\textrm{s}=0.34~\textrm{km~s}^{-1}$ for model E2. 
This means that both ejected clumps move supersonically, but the clump in model E2 moves with a higher Mach number through a denser environment than the ejected clump in model~E1. As a result, the ejected clump in model E2 fully disintegrates at $t=38$~kyr, leaving behind a bow-shock-like structured tail. On the contrary, the ejected clump in model E1 survives the ejection and continues its outward journey until it leaves the computational domain. The tail structures of both ejected clumps show a qualitatively similar bow-shock-like shape with a dense perimeter and a rarefied inner part. Such a shape is a result of supersonic motion of the ejected clumps, which create a bow shock in the ambient gas as they move outward.  The bow-shock morphology may be artificially enhanced in our thin-disk simulations, but observations of face-on and moderately inclined bow shocks can make a similar effect due to projection effects.

%%%%%%%%%%%%%%%%%  MODEL E3  %%%%%%%%%%%%%%%

\begin{table*}
\center
\caption{\label{tab:2}Parameters of ejected clumps}
\begin{tabular}{cccccccccccc}
\hline 
\hline 
Clump & $t$ & $M_{\mathrm{cl}}$ & $T_{\mathrm{cl}}$ & $v_{\mathrm{cl}}$ & $v_{\mathrm{esc}}$ & $a_{\mathrm{max}}$ & $M_{\mathrm{gr.dust}}$ & $M_{\mathrm{sm.dust}}$ & $E_{\mathrm{rot}}/E_{\mathrm{grav}}$ & $\zeta_{\mathrm{d2g}}$ \tabularnewline

 & [kyr] & [$M_{\mathrm{Jup}}$] & [K] & [km s$\mathrm{^{-1}}$] & [km s$\mathrm{^{-1}}$] & [$\mu$m] & [$M_{\mathrm{\oplus}}$] & [$M_{\mathrm{\oplus}}$] & & \tabularnewline
\hline 
F1 & 44  & 7.5 & 54 & 1.93 & 0.80 & 6.99 & 9.22 & 14.62  & 0.029 & 0.010\tabularnewline

F2 & 49  & 3.4 & 17 & 1.52 & 0.83 & 4.23 & 2.86 & 6.99  & 0.294 & 0.009\tabularnewline

F3 & 58  & 32.6 & 107 & 1.43 & 0.88 & 1250 & 31.14 & 81.04 & 0.526 & 0.011 \tabularnewline
\hline 
\end{tabular}
\center{ \textbf{Notes.} Time $t$ when parameters of the clumps were calculated, $T_{\mathrm{cl}}$ and $M_{\mathrm{cl}}$ are the central temperature and the total gas mass of the ejected clump, $v_{\mathrm{cl}}$ is the velocity of the clump, $v_{\mathrm{esc}}$ is the escape velocity of the system, $a_{\mathrm{max}}$ is the maximum radius of the dust particles inside the clump, $M_{\mathrm{gr.dust}}$ and $M_{\mathrm{sm.dust}}$ are the masses of grown and small dust particles in the clump, respectively, $E_{\mathrm{rot}}/E_{\mathrm{grav}}$ is the ratio of rotational to gravitational energy of the clump, and $\zeta_{\mathrm{d2g}}$ is the total dust to gas mass ratio averaged over the clump extent.}
\end{table*}

\begin{figure}
\begin{centering}
\includegraphics[width=\columnwidth]{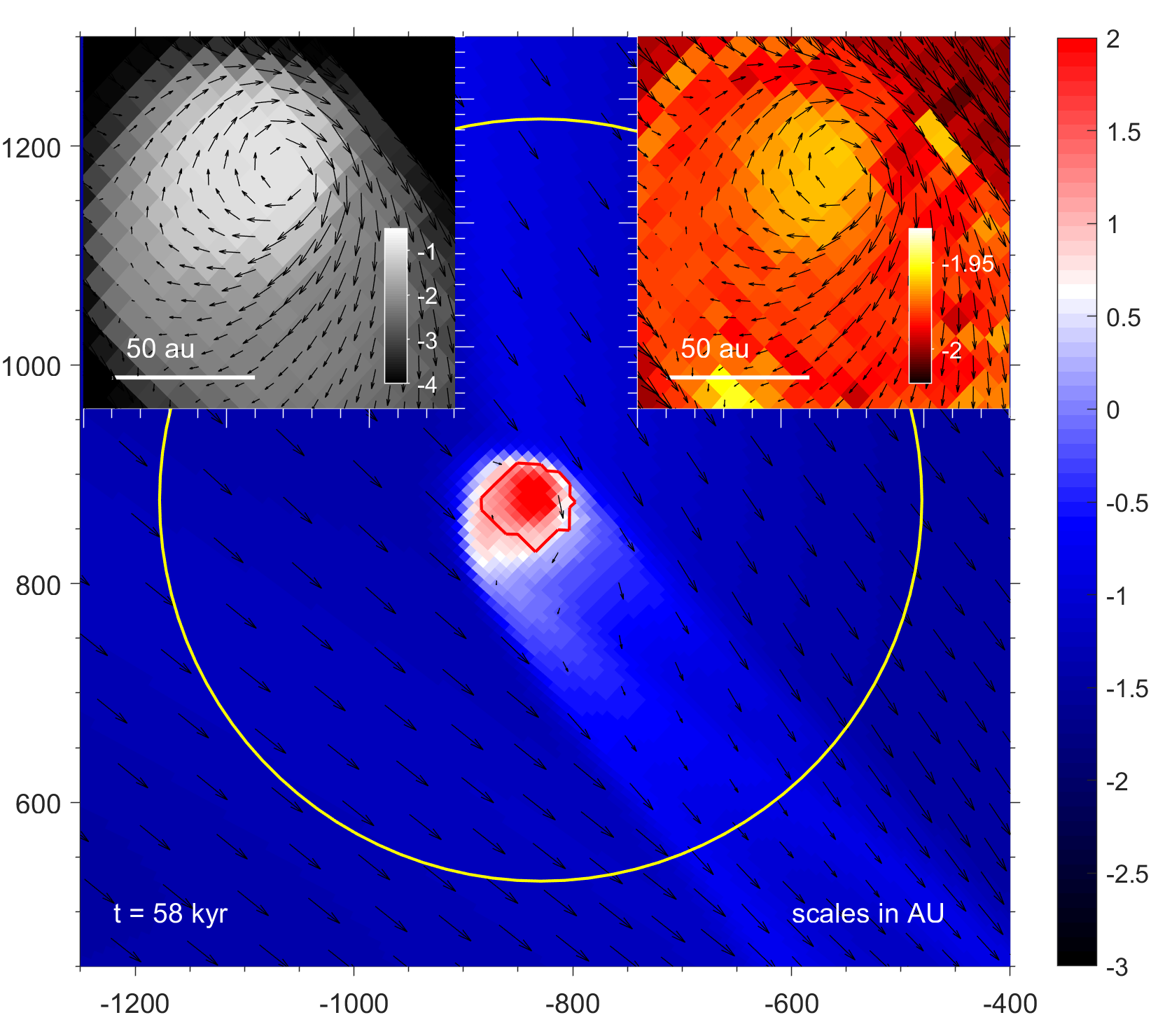}
\par\end{centering}
\caption{\label{fig:3} Gas surface density map in the vicinity of clump F3 at $t=58$~kyr. The color bar is in log~$\rm{g \ cm^{-2}}$. The yellow line shows the Hill radius of the ejected clump. The red contour line outlines the shape of the clump. The insets in the top-right and top-left corners show, respectively, the dust to gas mass ratio
%surface density (in log~$\rm{g \ cm^{-2}}$) 
and the maximum grown dust size (in log~cm) around the clump. The black arrows show the gas velocity field in the frame of reference of the clump.}
\end{figure}

We note that the clumps in our thin-disk simulations are necessarily ejected in the disk midplane. This may not be far from reality because the total angular momentum of the disk is mostly composed of the $z$-component, unless strong disk warps are present. As a consequence, the $z$-component will also dominate the total angular momentum of the clumps that form in the disk and the ejection will be predominantly in the disk plane.  In principle, clumps on highly eccentric orbits can turn into clumps on highly inclined orbits through the Kozai-Lubov effect, leading to possible off-the-plane ejections. However, we rarely see clumps on highly eccentric orbits because high eccentricity in the early stages of disk evolution is damped through the interaction of the clumps with the gas disk.

During the early evolution of model E3, multiple clump ejections over a time period of several tens of kyr were observed. Here we study in detail three different ejection events. Figure~\ref{fig:2} presents the gas surface density maps in model E3 for different time instances covering 20~kyr of disk evolution.   
%The Figure covers 20~kyrs of disk evolution during which three clump ejection events are observed. 
Each ejected clump is marked with the yellow arrow and labeled as F1, F2, and F3 based on the chronological order of ejections.
All the ejected clumps form bow-shock-shaped tail-like structures. Because of frequent multiple ejections, several tail-like structures in model E3 can co-exist, making their observational interpretation difficult \protect\footnote{The animation of clump ejections in model~E3 associated with Fig.~\ref{fig:2} is available at http://astro.sfedu.ru/animations/ejection.mp4.}. 

% {\textbf{The animation of clump ejections in model~E3 associated with Fig.~\ref{fig:2} is available at \protect\url{http://astro.sfedu.ru/animations/ejection.mp4}

The parameters of ejected clumps at a given time instance $t$ are presented in Table~\ref{tab:2}. All the ejected clumps have the radial velocity $v_{\rm cl}$ much higher than the escape velocity of the system $v_{\rm esc}$, meaning that the clumps will supposedly leave the system and become free-floating objects. The ejected clumps are characterized by a variety of gas masses. Clumps F1 and F2 have total gas masses of 7.5 and 3.4 $M_{\rm Jup}$, respectively, while clump F3 has a gas mass of 32.6 $M_{\rm Jup}$. These clumps also carry a significant dust content in the form of small dust grains, but also grown dust. The maximum size of grown dust can vary from several microns to millimeters for the most massive clump.
If these clumps survive ejection (the lowest mass clump, F2, may not), they may form free-floating giant planets and brown dwarfs with a near-solar metallicity as indicated by the mean dust to gas mass ratio $\zeta_{\rm d2g}\approx$~0.009-0.011 (see Table~\ref{tab:2}).
%However, as was shown by \citet{2018VorobyovElbakyan} only small part of the clumps eventually ends up forming planetary cores, while the main part of the clump is dissipating and/or forming a disk around central object. The total dust mass inside the all ejected clumps relates to gas as about 1:100. Thus, the ejection event could enrich the interstellar medium with dust. The mass of dust bearing by each ejected clump is presented in Table~\ref{tab:2}.

%%%%%%%%%%%%%%% ZOOM-IN %%%%%%%%%%%%%%%%%%

To obtain a more detailed understanding of the internal structure of the ejected clumps, we show in Figure~\ref{fig:3} the gas surface density map around clump F3 at $t=58$~kyr. Two insets in the top-left and top-right corners present the close-up views on the maximum dust size (in$\rm~log~cm$) and the total dust to gas mass ratio (in log units), respectively. The black arrows show the gas velocity field in the frame of reference of the center of the clump. 
The red contour line outlines the shape of the clump as defined by the clump-tracking method described in detail in \citet{2018VorobyovElbakyan}.
The yellow circle shows the Hill radius of F3 clump calculated as 
\begin{equation}
R_{\mathrm{H}}=r_{\mathrm{cl}}\left(\frac{M_{\mathrm{cl}}}{3(M_{*}+M_{\mathrm{cl}})}\right)^{1/3},
\end{equation}
where $M_{\mathrm{cl}}$ is the mass of the
clump confined within the red contour line, $M_\ast$ is the mass of the central star, and $r_{\rm cl}$ is the radial distance to the clump from the star. 
Clearly, the Hill radius captures the entire clump and
part of the tail, although the velocity field in the tail is indicative of a gradual loss of the swept-up material.
Clearly, the gas inside the clump is characterized with a strong clockwise rotation around the center of the clump, while the parental disk from which the clump was ejected rotates counter-clockwise. This implies that the ejected clump experienced multiple merging events in the disk before being ejected, which changed its direction of rotation.  
%MORE DETAILS ON REASONS? The grown dust effectively drifts inwards inside the ejected clump and is mainly concentrated in the central part of the clump\citep{2019VorobyovElbakyan}. 
Dust in the clump interior grows up to 1 mm in radius, which makes such a clump detectable with ALMA. We note that dust in the other two clumps, F1 and F2, grows only up to a size of several microns. We also note that the total dust to gas mass ratio in the center of F3 is 0.011 (not to be confused with the averaged over the clump $\zeta_{\rm d2g}=0.009$ in Table~\ref{tab:2}), slightly larger than the canonical 1:100 value, indicating that grown dust drifts towards the clump center (a natural pressure maximum).

%he ejected clump moves with super-sonic speeds through the sparse environment that has surface density $\Sigma\approx0.02 \ \rm g \ cm^{-2}$ and temperature $T\approx16$~K, thereby creating a shock wave in front of itself.

%The rotational to gravitational energy is ???. Centrifugal radii of the clumps. Temperatures in their center never reach 2000 K needed for the molecular hydrogen dissociation and gravitational collapse initialization.  

\begin{figure}
\begin{centering}
\includegraphics[width=1\columnwidth]{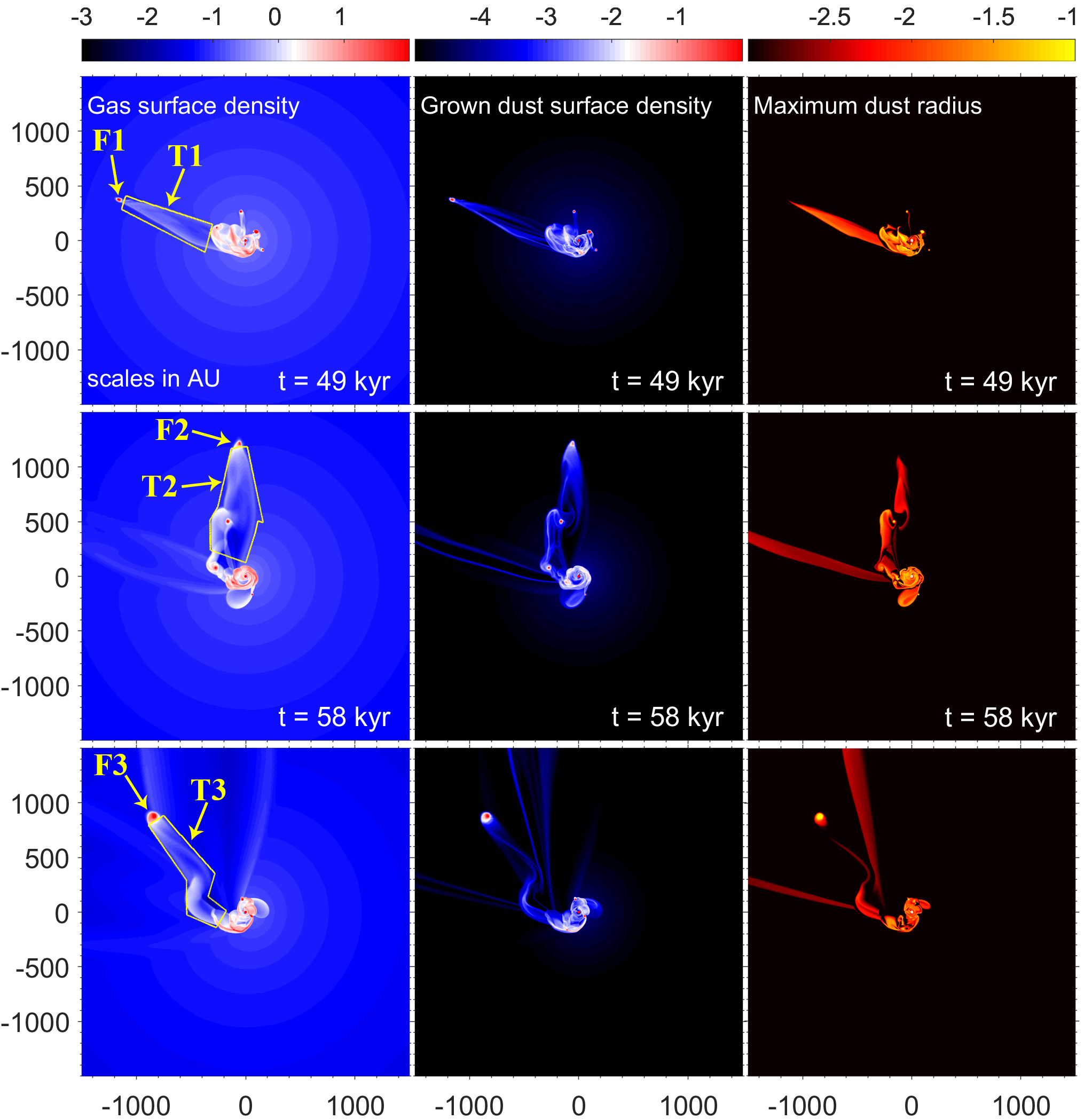}
\par\end{centering}
\caption{\label{fig:4} Gas surface density (left column), grown dust surface density (central column), and maximum  dust size (right column) maps in the inner $3000\times3000$~au$^2$ box for three ejection events in model E3. The color bars are shown in log scale in $\rm{g \ cm^{-2}}$ and in cm units. The yellow contour lines in the left column outline the clump-induced tail-like structures . }
\end{figure}

%%%%%%%%%%% TAIL STRUCTURES %%%%%%%%%%%

All the ejected clumps in model E3 create distinct tail-like structures extending for hundreds or even thousands of au. Below we present the detailed analysis of these structures. Each row in Figure~\ref{fig:4} shows the gas surface density (left column), grown dust surface density (middle column), and maximum size of grown dust (right column) in the disk of model E3 at different time instances. The time instances were chosen to show the disk when the ejected clumps, marked as F1, F2, and F3, reach a radial distance of $r\approx1200$~au from the star. The properties of these clumps are listed in Table~\ref{tab:2}. We note that the ejections occur at 40-50~kyr after the formation of the central star, when the disk is still embedded and young.

To measure the mass of gas and dust inside the tail-like structures, we analyze the segments of the disk outlined with the yellow contour lines in the left column of Figure~\ref{fig:4}. It is assumed that all material in the segments belongs to the tail-like structures. We note that such an approximation can slightly overestimate the gas  mass in the tail-like structures, while the mass of grown dust, which is dragged by the clump from the disk and is almost absent in the circumdisk environment, will not be affected.  The gas mass, along with the grown, small and total dust masses inside the tail-like structures, are presented in Table~\ref{table3}. 

\begin{figure*}
\begin{centering}
\includegraphics[width=2\columnwidth]{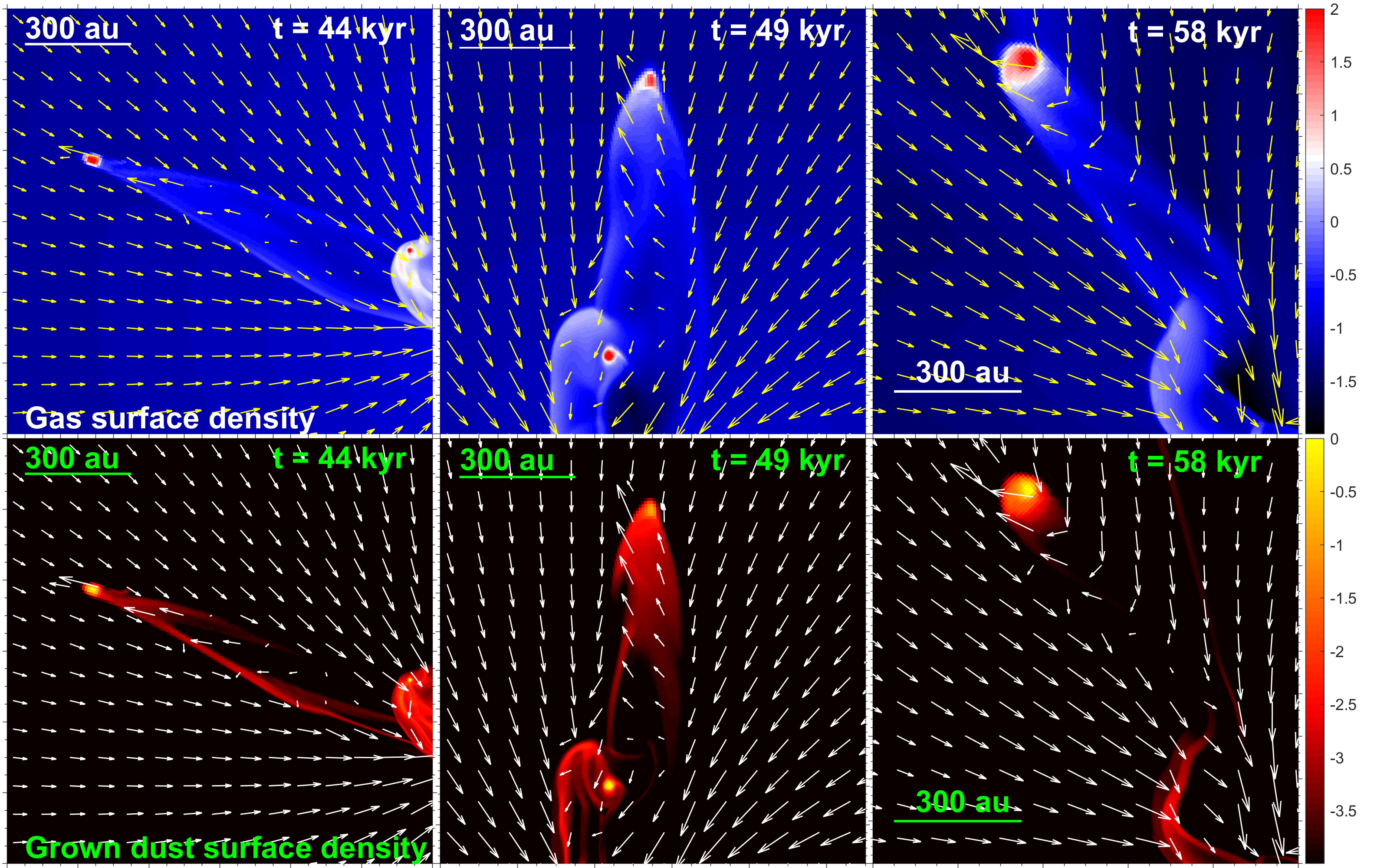}
\par\end{centering}
\caption{\label{fig:5} {\bf Top row. } Gas velocity field superimposed on the gas surface density distribution in model~E3 illustrating the gas flow in and around the clump-triggered tails. {\bf Bottom row.} Similar to the top row, but for grown dust. The scale bars are in $\log$ g~cm$^{-2}$. The arrows of maximum length correspond to 2.96~km~s$^{-1}$.}
\end{figure*}

The gas and dust content of the tails varies by a factor of three and lies in the 3.5--10.5~$M_{\rm Jup}$ and 10.6--31.1~$M_\oplus$ limits, respectively. The gas mass of F1 and F3  is smaller than the mass of the corresponding ejected clumps, while in the case of F2 the gas mass in the tail is much higher than the mass of the ejected clump. The total dust to gas mass ratios do not differ notably from the canonical 1:100 value.
%The gas and dust content of the tails is rather similar and varies only by a maximum factor of two. 
%The gas mass in the tails lies in the 3.5-10.4~$M_{\rm Jup}$ limits, which is comparable to or smaller than the mass in the ejected clumps. 
We note that clump F3 is quite massive and easily survives the ejection, losing only small part of its envelope.
Interestingly, the tails are dominated by small sub-micron dust.   On the other hand, the tail-like features look sharper in grown dust than in gas (and in small dust that follows gas due to strong frictional coupling). This is because the circumdisk environment (infalling envelope) lacks grown dust but is abundant in gas and small dust. 
%Therefore, it is not clear a priory which observation technique to use: scattered light, molecular tracers, or dust continuum. 
A maximum dust size of 0.1-0.37~mm suggests that these tails may  also be detected in the shortest wavelengths bands of ALMA.

The observational detection of the peculiar bow-shock-type tails can serve as a smoking gun for clump ejection (and disk fragmentation) even when the clump itself has dispersed or left the field of view. Unfortunately, each ejection event is a short-lived phenomenon. As Fig.~\ref{fig:2} demonstrates, the tail-like structure created by individual clumps start smearing out already after 5~kyr from the onset of ejection. 
We note that the tail associated with FU~Ori has $\sim 1300$~au in length. Assuming the velocity of the clump on the order of 1.5 km~s$^{-1}$, its dynamical age is approximately 4 kyr, which is consistent with our models.
On the other hand, a disk can produce several ejections, which makes this phenomenon more likely to observe. Still, the integrated lifetime of the tail-structures (several tens of kyr) is only a small fraction of the initial gravitationally unstable stage of disk evolution (several hundreds of kyr).

%Ejected F1 and F2 clumps are much less massive than F3 clump, making them less stable (vulnerable?) against the shock wave (supersonic movement?). During the ejection F1 and F2 clumps lose the main part of their outer envelopes, including quite large amount of grown dust, thus enriching their tail-structures with grown dust particles. In contrast, ejected F3 clump is quite massive and easily survives the ejection, loosing only small part of its envelope. As a result, the tail-like structure of F2 clump has small amount of grown dust in it. The mass of grown dust inside the tail structure of F3 clump is 20 times less than inside the F3 clump itself, while for F1 and F2 clumps this ratio is 8.7 and 1.4, respectively.
%Maximum radius of grown dust in the tail structures is presented in the last column of Table~\ref{tab:3}.

Finally, in Figure~\ref{fig:5} we present the gas and grown dust velocity maps in model~E3 superimposed on the gas and grown dust surface density distributions (top and bottom rows, respectively). Several time instances were chosen to illustrate the character of gas (and hence small dust) and grown dust motion in the clump-triggered tails. Notable outward flows are evident in the tails, whereas the surrounding medium is generally described by infalling motion towards the disk plus star system. This peculiar  character  of gas/dust flow can in principle be used to distinguish the clump-induced tails from other structures, such as large-scale inflowing streams of gas/dust that may connect a protostellar disk with a turbulent star formation environment  \citep{2014Padoan}.

\begin{table}
\center
\caption{\label{table3} Parameters of clump-triggered tails}
\resizebox{\columnwidth}{!}{\begin{tabular}{cccccccc}
\hline 
\hline 
Tail & $t$ & $M_{\mathrm{tot}}$ & $M_{\mathrm{gr.dust}}$ &  $M_{\mathrm{sm.dust}}$ &  $M_{\mathrm{gas}}$ & $a_{\mathrm{max}}$ & $\zeta_{\mathrm{d2g}}$ \tabularnewline

 & [kyr] & [$M_{\mathrm{Jup}}$] & [$M_{\mathrm{\oplus}}$] & [$M_{\mathrm{\oplus}}$] & [$M_{\mathrm{Jup}}$] & [$\mu$m] & \tabularnewline
\hline 
T1 & 44.0  & 3.49 & 1.06 & 9.67  & 3.46  & 220 & 0.0098\tabularnewline
T2 & 48.8  & 10.52 & 3.62 & 27.52 & 10.42 & 370 & 0.0094\tabularnewline
T3 & 57.9  & 8.41 & 1.92 & 25.33 & 8.32  & 100 & 0.0103\tabularnewline
\hline 
\end{tabular}}
\center{ \textbf{Notes.}  Time $t$ when parameters of the tails were calculated, $M_{\mathrm{tot}}$ is the total mass of dust and gas in the tails, $M_{\rm gr,dust}$ and $M_{\rm sm,dust}$ are the masses of grown and small dust in the tails, $M_{\rm gas}$ is the mass of gas in the tailsб $a_{\rm max}$ is the maximum size of dust in the tails, and $\zeta_{\rm d2g}$ is the total dust to gas mass ratio.}
\end{table}

\begin{table*}
\center
\caption{\label{col_par}Model parameters}
\begin{tabular}{cccccccc}
 &  &  &  &      \tabularnewline
\hline 
\hline 
Model 	& ${M_{\rm disk, init}^{\rm trg} }$		& ${M_{\rm \ast, init}^{\rm intr}} $  	& ${M_{\rm disk, fin}^{\rm trg}} $ 		& ${M_{\rm \ast, fin}^{\rm intr}} $ 	& ${ M_{\rm disk, fin}^{\rm intr}} $ & $R_{\rm per}$	& Intruder				\tabularnewline
		& [${\rm M_{Jup}}$] 		&  [${\rm M_{\odot}}$] 	& [${\rm M_{Jup}}$] 			& [${\rm M_{\odot}}$] 	& [${\rm M_{Jup}}$] &	[au]	&					\tabularnewline
\hline 
C1 	& 43    & 0.04	& 45.32	& 0.041  & 0.129(0.005)    & 213 & brown dwarf           \tabularnewline
C2 	& 43	& 0.08	& 43.55	& 0.082	 & 1.28(0.02)      & 208 & transitional object 	\tabularnewline
C3 	& 43    & 0.2 	& 40.2	& 0.204  & 4.45 (0.05)	   & 193 & very low-mass star 	\tabularnewline
C4 	& 43	& 0.5	& 32.67 & 0.51	 & 5.43(0.1)	   & 159 & half-solar mass star 	\tabularnewline
C3r & 43	& 0.2	& 54.49 & 0.2001 & 0.122 (0.0007)  & 191 & very low-mass star 	\tabularnewline
\hline
\end{tabular}
\center{ \textbf{Notes.} ${ M_{\rm disk,init}^{\rm trg} }$ is the initial disk mass of the target,  ${ M_{\rm \ast, init}^{\rm intr} }$ is the initial
mass of the intruder, $M_{\rm disk,fin}^{\rm trg}$} is the final disk mass of the target,  $M_{\rm \ast,fin}^{\rm intr}$ is the final stellar mass of the intruder, $M_{\rm disk,fin}^{\rm intr}$ is the final disk mass of the intruder (the numbers in parentheses show the final total dust mass), and $R_{\rm per}$ is the intruder perihelion.
\end{table*}

\section{Tail-like structures produced by close encounters with a (sub-)stellar object}
\label{flybys}

The second mechanism for the formation of tail-like structures that we consider is a close encounter with another (sub-)stellar-mass object.  {\ This mechanism was studied in a number of papers including \citet{2003Pfalzner}, which focuses on the long-term effects of collisions,  and \citet{2019Cuello}, which is similar to our study, though we focus mostly on the properties of tail-like structures.} The initial setup for this problem differs from that considered in Section~\ref{ejection}, where we studied the tail-like structures produced by ejected clumps in young and massive gravitationally unstable disks. 
%Older and more evolved disks are unlikely to exhibit fragmentation \citep[however, see][]{2019VorobyovSkliarevskii} because of disk masses quickly decreasing with the disk age. 
In this section, we focus on evolved disks with an age of $\sim 1.0$~Myr, because close encounters with a young gravitationally unstable disk was studied in detail in \citet{2017VorobyovSteinrueck}.

The initial disk configuration was taken from our numerical hydrodynamics simulations of gravitoviscous disks presented in \citet{2019VorobyovSkliarevskii}. In that work, we considered the effect that the mass transport rate through the central smart cell (CSC) can have on the global evolution of the entire disk. Following the notation adopted in \citet{2019VorobyovSkliarevskii}, we assume that the mass accretion from the CSC on the star is a fraction $\xi$ of mass accretion from the disk to the CSC
\begin{equation}
\dot{M}_{*, \rm csc}=
\begin{cases}
\xi \dot{M}_{\rm disk} &\text{for $\dot{M}_{\rm disk} > $ 0},
\label{Sink:1}
\\ 
0 &\text{for $\dot{M}_{\rm disk} \leqslant $ 0}.
\end{cases}
\end{equation}
The limit of small $\xi  \approx 0$ corresponds to slow mass transport through the CSC, resulting in gradual mass accumulation in the inner disk regions. Slow mass transport through the CSC may be caused by the development of a dead zone there. The opposite limit of large $\xi \approx 1.0$ assumes fast mass transport, so that the matter that crosses the CSC--inner disk interface does not accumulate in the CSC, but is quickly delivered on the star. Physically, this means that mass transport mechanisms of similar efficiency operate in the inner disk and in the CSC. In this work, we adopt $\xi=0.95$. The opposite case will be considered in a follow-up study focusing on accretion bursts triggered by close encounters.

We note that resolving the inner disk region in global simulations using an explicit integrator is extremely computationally costly, because the converging grid lines on the polar grid (adopted for our simulations) makes the Courant condition on the timestep too stringent. Our solution is to cut out the inner 1 au  and replacing it with a single cell. A more sophisticated approach would be to coarsen the computational grid as it approaches the star, thus easing the Courant limitation. This approach has, however, some serious disadvantages, the main being the problem with a fast and accurate calculation of the gravitational potential on a very irregular grid. We use the FFT transform technique to quickly calculate the gravitational potential on the polar logarithmic grid \citep{1987Binney}. On a non-regular grid this will not anymore work, and much slower and less accurate techniques have to be applied \citep{2008Murphy}.  We stress that resolving the inner disk regions is indeed a very nontrivial problem. The introduction of the CSC with a free inflow-outflow boundary condition allowed us, at the very least, to get rid of the artificial density drop near the inner boundary, which may develop if the sink acts only as a one-way valve, allowing for matter to flow from the disk to the sink, but not vice versa \citep{2019VorobyovSkliarevskii}.

\begin{figure*}
\begin{centering}
\includegraphics[width=1\textwidth]{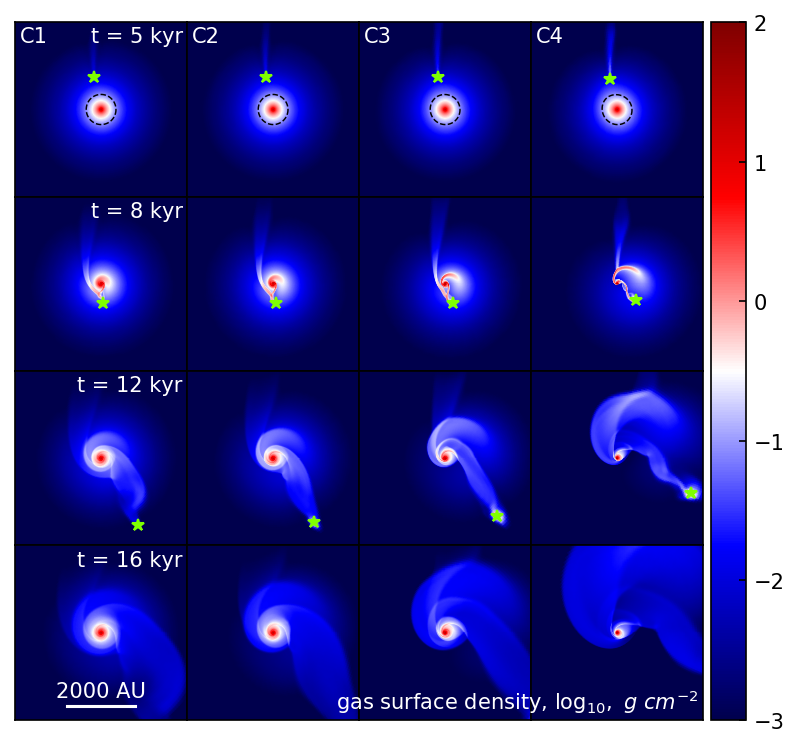}
\par\end{centering}
\caption{\label{gas_model3}  Gas surface density maps in the inner $5000 \times 5000$ au$^2$  box shown at four distinctive time instances (from top to bottom row) during a close encounter with an intruder of different mass (from left to right column): 0.04~$M_{\odot}$, 0.08~$M_{\odot}$,  0.2~$M_{\odot}$, and 0.5~$M_{\odot}$. The time is counted from the launch of the intruder. The current position of the intruder is indicated with the star symbols. The dashed circles in the top row correspond to $\Sigma_{\rm g}=0.1$~g~cm$^{-2}$ and thus roughly outline the outer disk extent.  The scale bar is in log~$\rm{g \ cm^{-2}}$.}
\end{figure*}

In the model of close encounters presented in \citet{2017VorobyovSteinrueck} only mass accretion on the target was taken into account and mass accretion on the intruder was neglected.
In this work, we also take mass accretion on the intruder into account. In addition, we calculate the total luminosity of the intruder and consider its input in the total energy budget of the system. More specifically, we track the position of the intruder on the numerical grid and determine the grid cell where the intruder currently resides. We then calculate the net mass increment per time step in this grid cell (both gas and dust) and assume that this additional mass is accreted on the intruder at every time step.  This method assumes that the increase in gas and dust mass is caused exclusively by the gravitational pull of the intruder, which is not a bad assumption considering that the disk of the target before the collision was nearly in a steady-state with a mass transport rate through the disk as low as a~few~$\times 10^{-9}~M_\odot$~yr$^{-1}$.

The total luminosity of the intruder was calculated as the sum of accretion and photospheric luminosities. The former is defined as
\begin{equation}
L_{\rm accr}^{\rm intr}= {1\over 2}{G M_{\rm intr} \dot{M}_{\rm intr} \over R_{\rm intr}},
\end{equation}
where $M_{\rm intr}$ is the intruder mass, $R_{\rm intr}$ the intruder radius, and $\dot{M}$ the mass accretion rate on the intruder. 
The photospheric luminosity of the intruder $L_{\rm phot}^{\rm intr}$ was calculated using the stellar evolution tracks of \citet{2008YorkeBodenheimer} for  the known mass of the intruder and its fixed age of 1~Myr.  The total luminosity of the intruder is further used to calculate the radiation flux (energy per unit time per unit surface
area)  absorbed by the surrounding material at radial distance  $r_{\rm intr}$ from the intruder
\begin{equation}
    F_{\rm irr}^{\rm intr} = {L_{\rm accre}^{\rm intr} + L_{\rm phot}^{\rm intr} \over 4 \pi (r_{\rm intr}^2 +\epsilon^2) } \cos{\gamma_{\rm irr}},
\end{equation}
where $\epsilon$ is a smoothing factor set equal to the size of the grid cell where the intruder currently resides and $\cos \gamma_{\rm irr}$ is the cosine of the incidence angle of radiation set equal to 0.025.

In this section, we consider five models, the initial 
parameters of which are listed in Table~\ref{col_par}. In particular, we consider four intruders with different initial masses: $M_{\rm \ast,intr}$=0.04, 0.08, 0.2, and 0.5~$M_\odot$, which correspond to a brown dwarf, a transitional object straddling the brown dwarf--star boundary, a very low mass star, and a star with mass similar to that of the target. The mass of the target star is equal to 0.41~$M_\odot$ and its value changes insignificantly during the simulations. All intruders are initially positioned at 3000 au to the north from the target. The initial radial velocity is the same and is equal to -2.5~km~s$^{-1}$. The azimuthal component is equal to 0.2 km~s$^{-1}$ for the models with prograde collisions (C1, C2, C3, and C4) and -0.2~km~s$^{-1}$ for the model with a retrograde collision (C3r).
All close encounters are in the plane of the disk of the target,  which is explained by the limitations of our thin-disk models. Off-plane encounters may introduce another degree of complexity in the morphology of tail-like structures and will hopefully be the subject of future studies.
The resulting perihelion distances range between 159~au and 213~au,
which is sufficient for the intruder to graze the outer disk regions and at the same time is not too small to be unlikely from the statistical arguments \citep[see, e.g.,][]{2010ForganRice}. 

\begin{figure*}
\begin{centering}
\includegraphics[width=2\columnwidth]{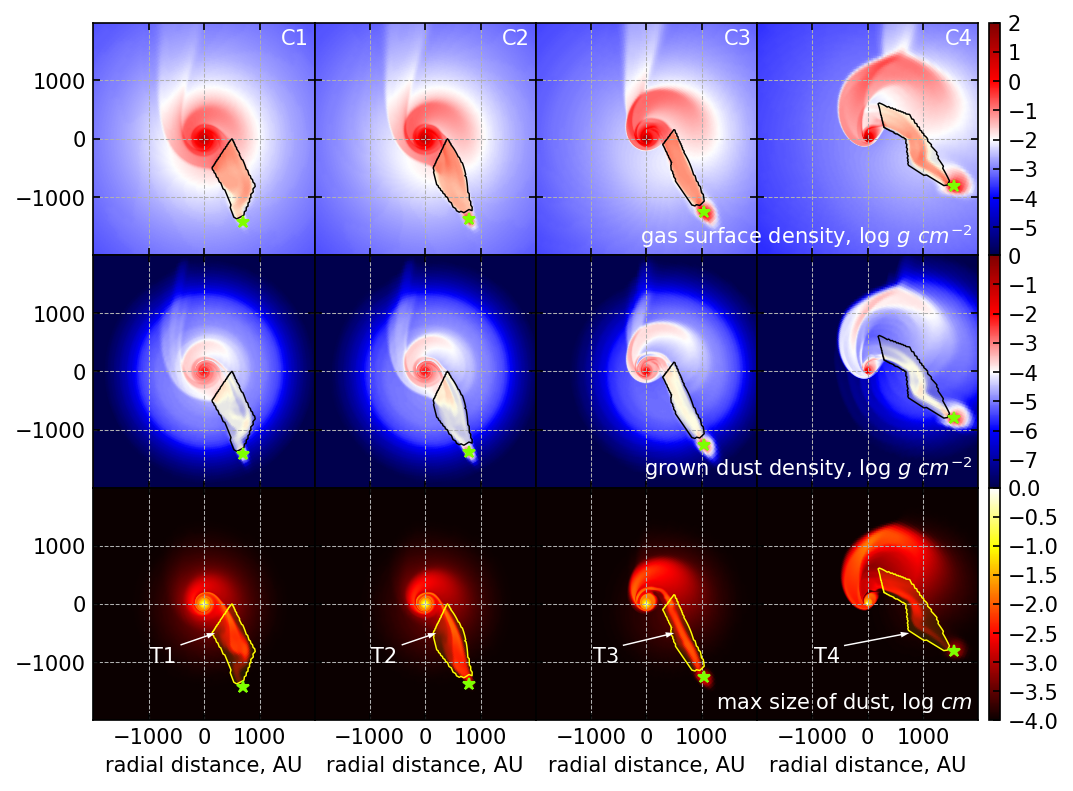}
\par\end{centering}
\caption{\label{post_col} Spatial distribution of gas surface density (top row), grown dust surface density (middle row), and maximal dust grain size (bottom row) shown for four models: \textbf{C1} (left column), \textbf{C2} (2nd column), \textbf{C3} (3rd column) and \textbf{C4} (right column). Tails are shown at 10.3 kyr after the launch of the intruder. The post-collisional tail structures are marked with the black and yellow contour lines. The star symbols indicate the position of the intruder. The gas and dust surface densities are in log~$\rm{g \ cm^{-2}}$, the maximum dust size is in log~cm.}
\end{figure*}

\subsection{Prograde encounters}

Figure~\ref{gas_model3} presents the gas surface density maps in models C1, C2, C3, and C4 corresponding to four distinctive phases of the collision: an approaching intruder (top row), a close interaction (second row), a receding intruder (third row), and a post-collisional system when the intruder has left the field of view (bottom row). The intruder is marked with the star symbol. The target star is in the coordinate center.  The dashed circles in the top row correspond to $\Sigma_{\rm g}=0.1$~g~cm$^{-2}$. This value is usually invoked to set the disk outer edge. In our case the disk lacks a sharp outer edge because of viscous spreading. On the other hand, the gas mass located outside of the dashed circle is only 15\% of the total gas mass contained within the shown $5000 \times 5000$~au$^2$ area. The dashed circles can therefore be regarded as an approximate size of the gas disk.

Upon close inspection, we have identified three types of tail-like structures triggered by the close approach with the intruder: a pre-collisional tail, a spiral tail, a post-collisional tail.  As can be expected, the strongest perturbation is created by the most massive intruder \protect\footnote{\textbf The animation of the encounter in model C4 associated with Fig.~\ref{gas_model3} is available at http://astro.sfedu.ru/animations/encounter.mp4.}. The tail-like structures in this case are also the sharpest, in contrast to more diffuse structures produced by low-mass intruders. Below, we consider each type of the tails in more detail.

Figure~\ref{post_col} shows the spatial distribution of gas, grown dust, and maximum grain radius (top, middle, and bottom rows, respectively) in models~C1, C2, C3, and C4 (from left to right). Here we focus on the phase after the closest approach when the intruder starts receding the target. The black and yellow lines outline the positions of the post-collisional tails. These tails are created by the gravitational pull of the intruder, which captures and drags along part of the disk material of the target.
Clearly, the tails appear sharper and narrower as the mass of the intruder increases. The tails stand out clearly against the low density environment both in gas (also small dust) and grown dust. It is also evident that sub-solar-mass intruders can capture appreciable gaseous and dusty disks around them (see Table~\ref{col_par}). For instance, the 0.5~$M_{\rm \odot}$ intruder in model~C4 can acquire a circumintruder disk of 5.43~$M_{\rm Jup}$ after a prograde encounter, of which 0.1~$M_{\rm Jup}$ is in the form of dust. This means that the captured disk is overabundant in dust with a dust to gas ratio of $\approx 0.019$ as compared to the canonical 1:100 value.

\begin{table}
\center
\caption{\label{tab:3} Collision-triggered tail parameters}
\resizebox{\columnwidth}{!}{\begin{tabular}{ccccccccc}
\hline 
\hline 
Model & Tail & $type$ & ${\rm M_{tot}}$ & ${\rm M_{gr.dust}}$ &  ${\rm M_{sm.dust}}$ &  ${\rm M_{gas}}$ & $a_{\mathrm{max}}$ & $\zeta_{\mathrm{d2g}}$  \tabularnewline

 &  &  & [$M_{\mathrm{Jup}}$] & [$M_{\mathrm{\oplus}}$] & [$M_{\mathrm{\oplus}}$] & [$M_{\mathrm{Jup}}$] & [$\mu$m] & \tabularnewline
\hline 
T1 & C1  & post-col & 3.57  & 1.384 & 6.92  & 3.55  & 109 & 0.0073 \tabularnewline
T2 & C2  & post-col & 3.83  & 1.386 & 6.97  & 3.81  & 118 & 0.0069 \tabularnewline
T3 & C3  & post-col & 3.6   & 1.252 & 6.47  & 3.57  & 105 & 0.0068 \tabularnewline
T4 & C4  & post-col & 3.18  & 1.129 & 5.85  & 3.16  &  38 & 0.0069 \tabularnewline
T5 & C3  & spiral   & 6.71  & 2.471 & 12.03 & 6.67  & 147 & 0.0068 \tabularnewline
T6 & C4  & spiral   & 11.88 & 3.911 & 21.64 & 11.8  & 131 & 0.0068 \tabularnewline
T7 & C1  & pre-col  & 1.73  & 0.618 & 3.08  & 1.72  &  21 & 0.0067 \tabularnewline
T8 & C4  & pre-col  & 0.86  & 0.196 & 1.56  & 0.85  & 4.2 & 0.0064 \tabularnewline
T9 & C3r & pre-col  & 3.02  & 1.068 & 5.41  & 3.00  &  40 & 0.0068  \tabularnewline
T10& C3r & post-col & 3.87  & 1.460 & 6.89  & 3.84  &  24 & 0.0068 \tabularnewline
%T9 & C3r & pre-col  & 1.256 & 0.423 & 2.23  & 1.24  &  28 &  \tabularnewline ???
%T10& C3r & post-col & 0.1   & 0.015 & 0.18  & 0.097 & 1.4 &  \tabularnewline ???
\hline 
\end{tabular}}
%\center{ \textbf{Notes.}   $M_{\mathrm{tot.dust}}$ is the total mass of dust in the tail structure created by the ejected clump.}
\end{table}

The parameters of the tails are provided in Table~\ref{tab:3}, where T1, T2, T3, and T4 are the post-collisional tail structures with the total masses from 3.18~$M_{\rm Jup}$ to 3.83~$M_{\rm Jup}$. The masses of dust (small and grown) within the tails are typically 0.6\%--0.7\% that of the gas mass, meaning that the tails are depleted in dust as compared to the canonical 1:100 value. Dust grows up to 118~$\mu$m in T2 and is smaller in other tails. Therefore, it might be possible to detect the post-collisional tail-like structures similar to T2 using sub-mm dust continuum observations (e.g., the ALMA bands 7-10). However, the T4 tail with a maximum dust size of just 38~$\mu$m is unlikely to be detected by ALMA.

\begin{figure}
\includegraphics[width=\columnwidth]{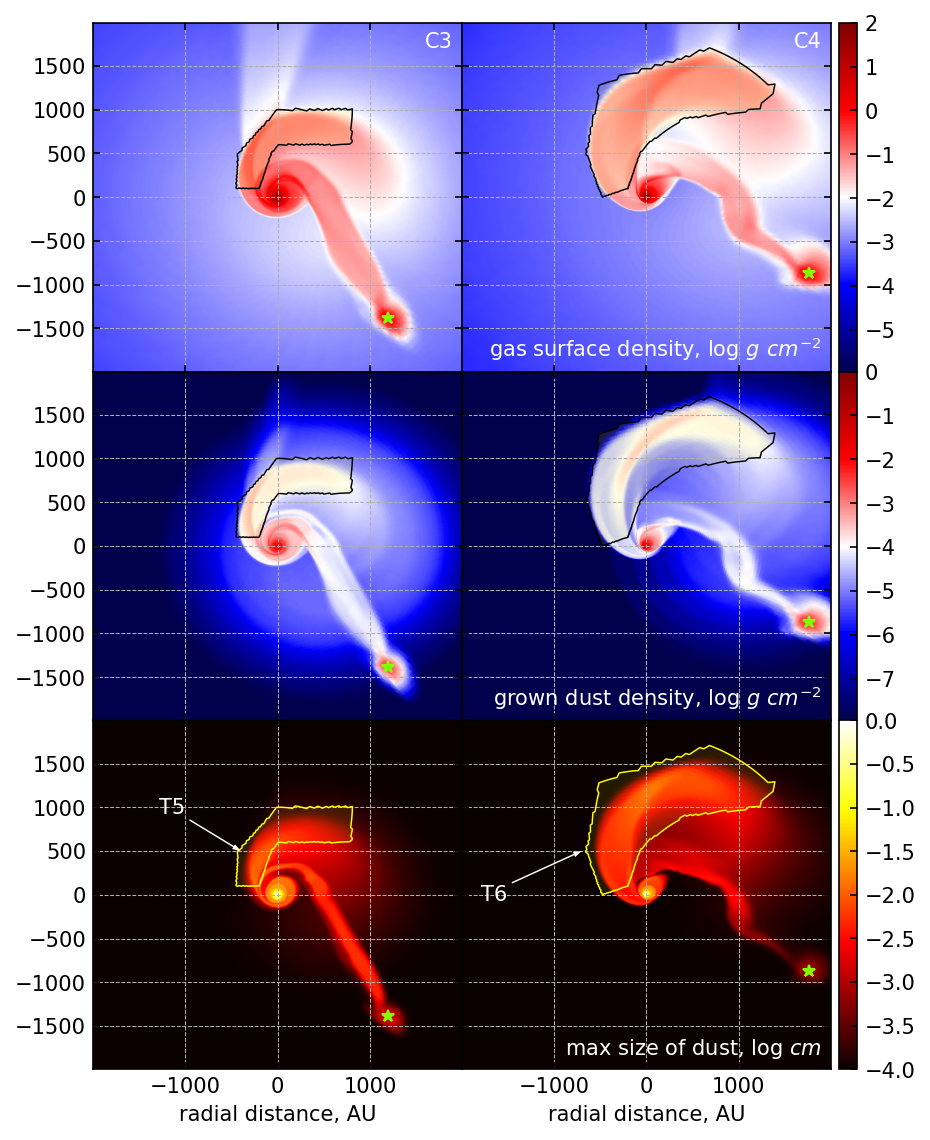}
\caption{\label{spi_lik} Spatial distributions of gas, grown dust, and maximum dust size (from top  to bottom row) in models C3 and C4 (left and right columns, respectively) at 10.8~kyr after the launch of the intruder, illustrating the formation of spiral tails outlined with the black and yellow contour lines. The gas and dust surface densities are in log~$\rm{g \ cm^{-2}}$, the maximum dust size is in log~cm.}
\end{figure}

The second type of tail-like structures -- the spiral tails -- are outlined with the black and yellow contour lines in Figure~\ref{spi_lik}. We consider only two models C3 and C4, for which this type of the tail-like structures is most pronounced. These tails are created by the gravitational pull of the intruder, which perturbs the orbits of gas and dust particles in the disk of the target as it flies by. Differential rotation in the target disk curves the resulting density wave into a spiral arm. 
Similar structures are readily formed in  numerical simulations of close encounters with an intruder star \citep[e.g.,][]{2010Thies}.  The spiral tails appear sharper in grown dust  rather than in gas, which could be explained by additional drift of grown dust toward the center of the spiral arms, which present natural pressure maxima in the surrounding environment.

The parameters of the spiral tails T5 and T6 are  listed in Table~\ref{tab:3}. The spiral tails are the most massive ones among all the considered tails. The total masses of T5 and T6 are 6.71~$M_{\rm Jup}$ and 11.88~$M_{\rm Jup}$, respectively, of which 14.5~$M_{\mathrm{\oplus}}$ and 25.55 $M_{\mathrm{\oplus}}$ are in the form of dust. The tails are depleted with dust with the total dust-to-gas mass ratios around 0.7\%. These tails are also characterized by the highest values of the maximal dust size, which are equal to 131~$\mu$m and 147~$\mu$m for the T5 and T6 tails, respectively. Taking conservative assumptions about the dust temperature (20~K), dust mass opacity at 345 GHz (1~cm$^2$~g$^{-1}$), and grown dust surface density of the tails ($10^{-4}$~g~cm$^{-2}$), the expected brightness temperature is $20 \times 1 \times 10^{-4} = 2$~mK.  With 1" angular resolution, default weather condition ($PWV\approx 1$~mm), and 40 minutes on-source integration we can achieve an RMS brightness temperature of 0.3 mK in Band 7 (354 GHz), meaning that the spiral-like tails may be detectable by ALMA in the (sub-)mm-waveband of dust continuum.  We note that these estimates and the choice of the ALMA band may be sensitive to the dust opacity model and to the free parameters of the dust growth model.

\begin{figure}
\includegraphics[width=\columnwidth]{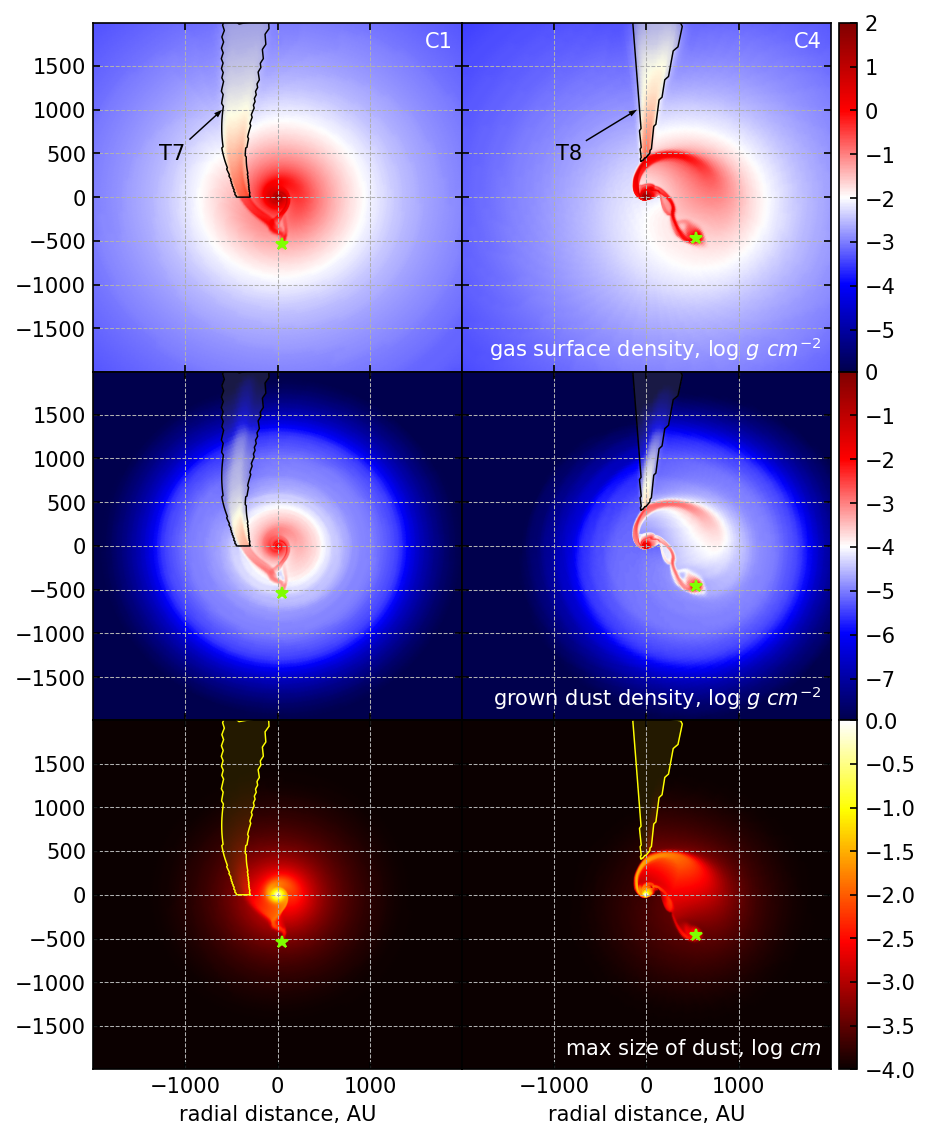}
\caption{\label{pre_col} Spatial distributions of gas, grown dust, and maximum dust size (from top  to bottom row) at 7.8 kyr after the launch of the intruder in models C1 and C4 (left and right columns, respectively), illustrating the formation of pre-collisional tails outlined with the black and yellow contour lines. The gas and dust surface densities are in log~$\rm{g \ cm^{-2}}$, the maximum dust size is in log~cm.}
\end{figure}

The black and yellow contour lines in Figure~\ref{pre_col} highlight the third type of tail-like structure found in our numerical simulations --  the pre-collisional tails -- which  are created by an intruder during its approach to the target star.  We show only models~C1 and C4 as the two limiting cases. The definitive feature of these tails is a rather small maximum size of dust grains, only a few microns. These tails are created by sweeping up gas and dust in the circumdisk low-density environment created by gradual viscous spreading of the outer disk, where dust has not grown appreciably compared to the interstellar dust spectrum.
%This means that these tails can be best detected with the scattered light or molecular tracers, but will be mostly undetected with dust continuum observations at, for example, ALMA. 
Table~\ref{tab:3} provides the characteristics of the T7, T8, and T9 pre-collisional tails. The masses of gas and small dust in these structures are the lowest ones. For example, the gas mass in T7 is 1.72~$M_{\rm Jup}$ and the mass of small dust is 3.08~$M_{\mathrm{\oplus}}$. The tails are slightly depleted with dust (the dust-to-gas mass ratios are 0.67\% and 0.64\% for T7 and T8, respectively) in comparison to the typical ISM value of 1\%. Moreover, dust grains are characterized by a rather small size (21 $\mu$m in T7 and 4.2 $\mu$m in T8), which may make their detection  problematic for the dust continuum observations with ALMA.

%The intruder sweeps up and drags behind the pristine material in the circumdisk environment, unprocessed by the disk.

\begin{figure*}
\begin{centering}
\includegraphics[width=2\columnwidth]{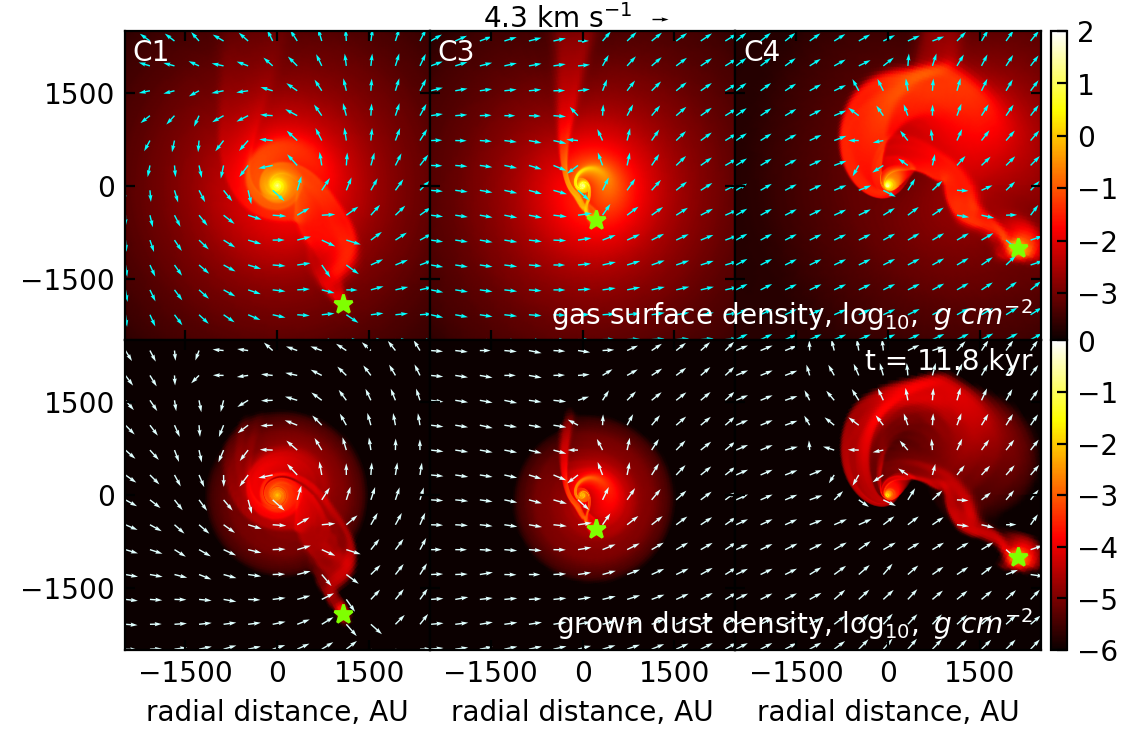}
\par\end{centering}
\caption{\label{velo} Surface density distributions of gas (top row) and grown dust (bottom row) with the superimposed velocity fields that illustrate distinctive flows in different types of tail-like structures: post-collisional tails (left column), pre-collisional tails (middle column),  and spiral tails (right column). The  cyan and white arrows indicate the velocity fields of gas and  grown dust, respectively. The arrow length is in the log scale. The arrows of maximum length correspond to 4.3~km~s$^{-1}$.
%Length of the arrows corresponds to the physical value of the velocity in logarithmic scale. 
The gas and dust color scale bars are in log~$\rm{g \ cm^{-2}}$.}
\end{figure*}

Figure~\ref{velo} presents the gas and grown dust velocity fields (top and bottom rows, respectively) superimposed on the corresponding surface density distributions in model~C1 (left column), model~C3 (middle column), and model~C4 (right column). These models were chosen to illustrate the character of gas and dust flows in post-collisional tails (left column), pre-collisional tails (middle column), and spiral tails (right column). The position of the intruder is marked with the star symbols. As expected, the gas and dust flow in pre-collisional and post-collisional tails follows that of the intruder trajectory -- in the former the flow is directed along the tail toward the disk of the target, while in the latter the flow is  oriented along the tail but away from the disk. On the other hand, the spiral tails are characterized by a peculiar velocity pattern that is directed away from the disk of the target but is simultaneously perpendicular to the tail. 
This flow pattern reflects an expansion motion, which is caused by the positive gravitational torque exerted on the spiral tail by the intruder as it flies by the disk of the target \citep[see][]{2017VorobyovSteinrueck}. As a result, the spiral tail gains angular momentum and expands. We note that a similar pattern is also seen in the inner parts of the post-collisional tail in model~C4.
We argue that the distinctive features of the gas and dust flows in the tails can be in principle used to distinguish their origin. Synthetic line emission maps of gas tracers (e.g., CO) are needed to justify our conclusion.

\begin{figure}
\includegraphics[width=\linewidth]{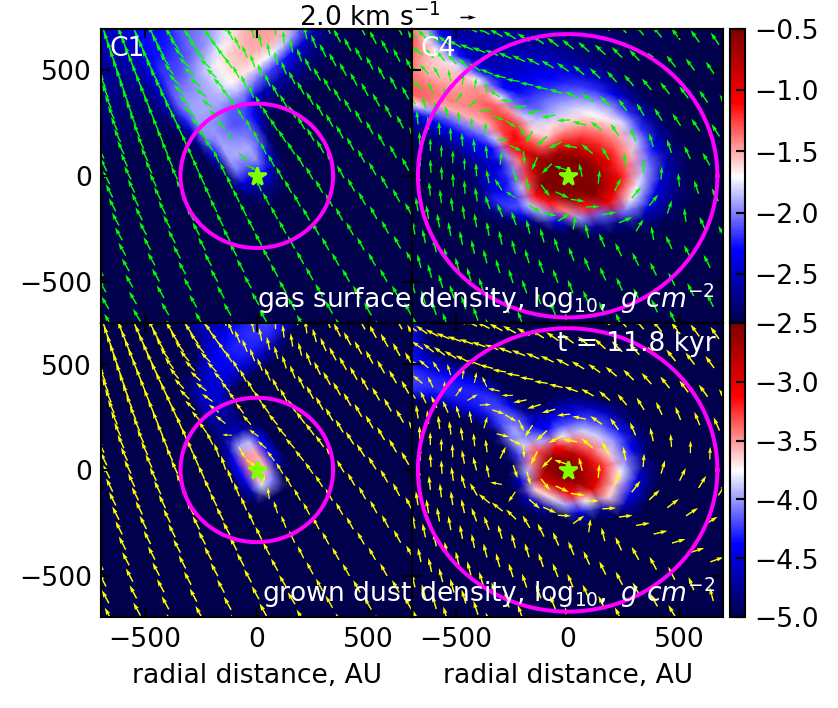}
\caption{\label{velo_zoom} Zoomed-in images focusing on the intruders in model~C1 ($M_{\rm \ast, intr}=0.04~M_\odot$, left column) and model C4 ($M_{\rm \ast,intr}=0.5~M_\odot$, right column), respectively. Shown are the gas (top row) and grown dust (bottom row) surface density distributions in the $1000 \times 1000$ au$^2$ area with the corresponding velocity maps calculated in the framework co-moving with the intruder. The gas and grown dust velocity fields are shown with the green and yellow arrows, respectively. The arrow length is in the log scale. The arrows of maximum length correspond to 2 km s$^{-1}$. The magenta circles correspond to one-half of the Hill radius of the intruder.}
\end{figure}

Finally, in Figure~\ref{velo_zoom} we zoom in on two intruders representing a brown dwarf of 0.04~$M_\odot$ (model C1, left column) and a star of  0.5$~M_\odot$ (model~C4, right column). The top/bottom rows show the gas/grown dust velocity fields superimposed on the gas/grown dust surface density distributions. The purple circles outline the Hill radii of the intruders. The velocity field is calculated in the frame of reference of the intruder. Clearly, the intruder star in model~C4 captures an appreciable disk with a mass of approximately 5.4~$M_{\rm Jup}$ (see Table~\ref{col_par}). The captured disk shows a distinctive rotational motion around the intruder in the same direction as the disk of the target. The average total dust to gas mass ratio in the captured disk is 0.019. However, the maximum dust size is only 5.8~$\mu$m because the intruder collects the matter from the outer disk regions of the target, where dust has not grown appreciably or has drifted towards the disk inner regions. 
This suggests that such circumintruder disks will remain almost undetectable in dust continuum with ALMA. 

The brown dwarf in model~C1 captures only a tiny disk. In fact, the captured material does not exhibit a rotational motion around the intruder, as would have been expected for a Keplerian disk.  Instead, the intruder drags behind some of the material from the target disk. The mass of the captured disk is only 0.129~$M_{\rm Jup}$ with an averaged total dust-to-gas mass ratio of 0.016. This implies that a low-mass intruder would be difficult to detect in gas or dust tracers and deep photometry may be needed to infer its presence.

In conclusion, we note that all types of the tail-like structures considered in this section lack the bow-shock-type morphology, which is typical of clump-triggered tails. One may think that this is because the evolved disks considered in the encounter models do not possess extended envelopes. However, collisions with young disk plus envelope systems were considered in \citet{2017VorobyovSteinrueck}. Somewhat surprisingly, we did not observe such a peculiar morphology in those simulations as well. The reason for this difference is uncertain and may be partly explained by different trajectories of the intruder and the clump -- the former flies along a curved path just slightly grazing the disk of the target while the latter flies almost radially outward through the contracting envelope.   We plan to investigate in  more detail this difference in the morphology of the tails in follow-up studies.
Here we only want to note that the  bow-shock-type structures can be indicative of a peculiar type of the tail-triggering mechanism -- clump ejection.

\begin{figure}
\includegraphics[width=\columnwidth]{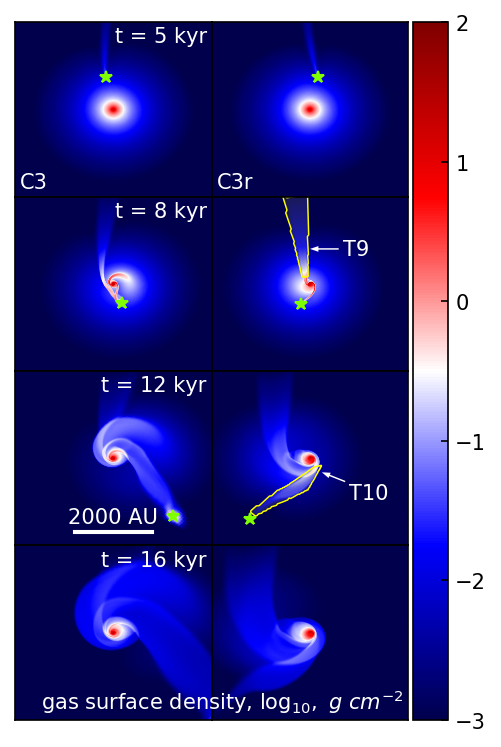}
\caption{\label{retr_gas} Prograde and retrograde collision in models C3 (left column) and C3r (right column) at several consecutive time instances. Shown are the gas surface densities in the inner $5000\times5000$~au$^2$ box. The position of the intruder is marked with the star symbols. The scale bar is in log~$\rm{g \ cm^{-2}}$.}
\end{figure}

\subsection{Retrograde encounter}

In this subsection, we compare the results of prograde and retrograde collisions. For this purpose, we launched exactly the same intruder as in model~C3, but with the opposite azimuthal velocity. The resulting gas surface density distributions at several time instances are shown in Figure~\ref{retr_gas}. In particular, the left and right columns correspond to the prograde model C3 and retrogade model C3r, respectively. While the pre-collisional tails look fairly similar, the post-collisional dynamics is remarkably different in the considered models. In particular, the retrograde intruder carries almost no disk around it, whereas the prograde intruder catches an appreciable disk (see also Fig.~\ref{velo_zoom}). Nevertheless, the post-collisional tails in both cases have similar gas and dust masses.  The second row also shows that the retrograde intruder does not trigger a spiral tail, however in the subsequent evolution the complicated gas flow distorts the circumstellar structure, so that the pre-collisional tail starts resembling the spiral one.

\begin{figure}
\includegraphics[width=\columnwidth]{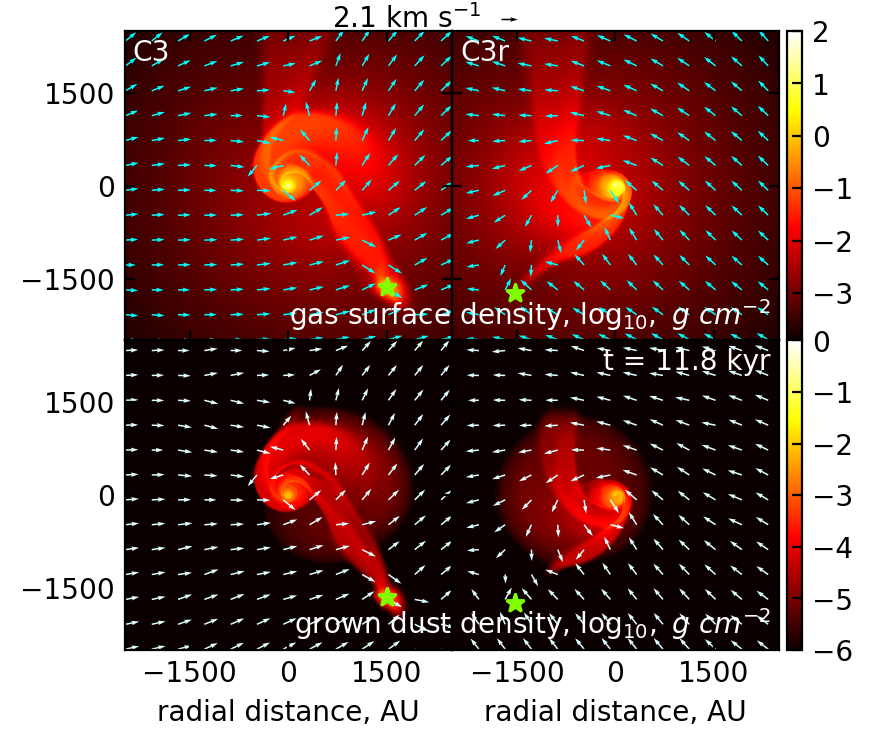}
\caption{\label{retr_vel} Gas and grown dust surface density distributions (top and bottom rows, respectively), together with superimposed velocity fields, in models~C3 (prograde collision, left column) and C3r (retrograde collision, right column). The gas and grown dust velocity fields are shown with the green and yellow arrows, respectively. The arrow
length is in the log scale. The arrows of maximum length correspond to 2.1~km~s$^{-1}$. The gas and dust color scale bars are in log~$\rm{g \ cm^{-2}}$. }
\end{figure}

In Figure~\ref{retr_vel} we present a comparison of the gas and grown dust velocity fields in models C3 and C3r. A distinctive feature of the post-collisional tail in the retrograde model is a converging flow of gas and dust toward the tail on both sides of the tail, while in the prograde model the velocity vectors are oriented toward the tail only on one side. In principle, this peculiar feature can be used to distinguish between prograde and retrograde collisions.

Finally, we note that the resulting morphology during the retrograde encounter at t=12~kyr  resembles the corresponding structures seen in the near-IR images of FU~Ori shown in \citet{2016LiuTakami} and \citet{2018Takami}. The retrograde encounter also explains why the disk seems to be centered on FU~Ori instead of the more massive component, FU~Ori~S, -- intruders on retrograde orbits do not capture appreciable circumintruder disks.

%\begin{figure}
%\includegraphics[width=\columnwidth]{figures/coll/model_3_quiver_zoom_retro.png}
%\caption{\label{retr_gas} }
%\end{figure}

\section{Conclusions and future prospects}
\label{conclude}

In this work we studied numerically the origin of the tail-like structures recently detected around disks of SU Aur, RW Aur, and several FU Orionis-type objects. We considered two scenarios: an ejection of a gaseous clump from a young disk that is prone to gravitational fragmentation and a close encounter of an intruder (sub-)stellar object with an evolved gravitationally stable disk. For this purpose, we used the FEOSAD numerical hydrodynamics code, which considers the mutual evolution of gas and dust, including gas-to-dust friction and dust growth \citep{2018VorobyovAkimkin,2019VorobyovSkliarevskii}. The close encounter was modelled in the plane of the target disk. We found the following.

- Gaseous and dusty clumps with total masses from a few to few tens of Jupiters can be ejected from young protostellar disks that are still embedded in the parental cloud cores. As these clumps transverse through the surrounding gaseous envelope, they create tail-like structures that are characterized by a  bow-shock-like shape with a dense perimeter and a rarefied inner part. Such a shape is a result of supersonic motion of the ejected clumps, which create a bow shock in the ambient gas as they move away from the disk. The length of these structures can reach a thousand of au or even more.  Such a peculiar bow-shock morphology was not found in other types of the tails.

- The gas mass in the clump-triggered tails ranges from 3.5~$M_{\rm Jup}$ to 10.4~$M_{\rm Jup}$, while the mass of dust can amount to 30~$M_{\rm \oplus}$, of which up to 3.6~$M_\oplus$ is in the form of grown dust with a maximum size of 0.37~mm. We speculate that these tails may be detected in the scattered light or dust continuum observations in the shortest wavebands of ALMA. 
Dust in the interiors of the ejected clumps grows to sizes of up to 1~mm. However,  clumps may be fully disintegrated during the ejection and the tail-like structures can serve as an indicator of clump ejection from the gravitationally fragmenting disk.

- Close encounters of a (sub-)stellar object with an evolved and gravitationaly stable disk can produce three different types of the tail-like structures: the pre-collisional tails, post-collisional tails, and spiral tails. These structures are better pronounced and more massive in the case of prograde encounters than in the case of retrograde ones.  Nevertheless, the best resemblance with the structure of the circumdisk environment in FU Ori was found for the case of retrograde encounters.

- The pre-collisional tails are formed when the intruder approaches the target disk and sweeps up matter in the low-density circumdisk environment created by a gradual viscous spreading of the outer disk regions. These types of the tails are least massive and are characterized by dust sizes that barely exceed a few microns.
%, which makes them potentially detectable only in the scattered light.

- The post-collisional tails are caused by the passage of the intruder through the target disk. The intruder tails gravitationally captures and drags behind part of the target disk. These tails are fairly massive (a few Jupiter masses of gas and several Earth masses of dust) and are characterized by maximum dust sizes up to 0.1~mm, making them potentially detectable in the scattered light and probably also millimeter dust continuum emission. Depending on the mass of the intruder and only in the case of prograde encounters, an appreciable disk with a well-defined rotational signatures can also be captured by the target. 

- The spiral tails are created by the gravitational perturbation exerted by the intruder on the disk of the target. Differential rotation curves the perturbation into a spiral arc and the positive gravitational torque exerted by the intruder causes the arc to gradually expand and leave the disk of the target \citep[see][]{2017VorobyovSteinrueck}.
Spiral tails are the most massive ones containing up to ten Jupiter masses in gas and tenths of Earth masses in dust with a maximum size of 0.1~mm, which makes this type of the tails the best candidates for detection in the scattered light and dust continuum.

%Massive intruders can capture the appreciable disk with a distinctive rotational motion. Retrograde collision leads to formation of significantly lower-massive tail-like structures. In this case dust growth is also less effective. 
%Tails formed during encounters can be observed by scattered light observations or (in case of post-collisional and spiral tails) sub-mm wavelength observations.

- The ejected clumps and the clump-triggered tails are characterized by the total dust to gas ratios that are similar to the canonical 1:100 value. The intruder-triggered tails are rather dust-poor with the dust to gas ratios lying in the 0.64\%--0.73\% limits.  Nevertheless, the total mass of dust in the tail-like structures lies in the 1.75--30.1~$M_\oplus$ range, which is higher than what was inferred for similar structures in SU~Aur, FU~Ori, and Z~CMa. The circumintruder disks, on the other hand, are dust-rich with the dust to gas ratios reaching 1.9\%.

- The considered tail-like structures are characterized by peculiar gas and dust flows in and around them, which may help to distinguished them from each other and from the surrounding circumdisk environment. 

What concerns the observability of the tails, grains of approximately 0.1--1 mm in size can be best traced by the (sub-)mm wavelength observations and the tail structures shown by our simulations can be examined by state-of-the-art facilities such as ALMA. For instance, the bow-shock-like structure with a dense perimeter and a rarefied inner part generated by  clump ejection can be observationally analyzed through a combination of multi-band observations that cover a range of mm/submm in wavelength and high angular resolution that spatially resolves the tail structure as demonstrated in SU Aur and RW Aur cases \citep{2019AkiyamaVorobyov,2018Rodriguez}. Multi-band observations can produce a so-called spectral index map or $\beta$-map, which is helpful in inferring the dust size distribution \citep{2019Pavlyuchenkov}. 

 Polarization observations also help identify grain sizes. \citet{2017Kataoka} estimated the grain sizes in a protoplanetary disk around HL~Tau through the polarization pattern taken in several wavelengths. If dust grains align in one direction or of uniform shape in the tail, grain size can be derived by examining the polarization vectors in the tail structure in several wavelengths.  Small dust grains with sizes of 0.1--1~$\mu$m  can be inferred  through molecular emission of for example CO in (sub-)mm wavelengths, as well as through the near infrared  scattered emission.  CO is generally optically thick and thus suitable for detecting a thin component in the tail, such as surface of the tail, which is helpful for visualizing  the tail’s outline.

 To further discuss the actual detection of the tails described above, we need to consider the effects of dust surface density, emissivity or scattering albedo, beam filling factor of the observations, and detection limits of the facilities. This would be the subject of our future work, similar to what has recently been done by \citet{2019Cuello}. Finally, we note that we have not considered the entire possible spectrum of impact parameters, intruder velocities, and disk parameters of the target. The intruder may also posses a disk and collisions with starless cloudlets are also feasible. This should also be the subject of future investigations. 

\section*{Acknowledgements}
We are thankful to the anonymous referee for constructive and thought-provoking comments that helped to improve the manuscript. This work was supported by the Russian Fund for Fundamental Research, Russian-Taiwanese project 19-52-52011 and MoST project 108-2923-M-001-006-MY3. H.B.L. is supported by the Ministry of Science and
Technology (MoST) of Taiwan, grant No. 108-2112-M-001-002-MY3. V.G.E. acknowledges the Swedish Institute for a travel grant allowing to visit  Lund University. The simulations were performed on the Vienna Scientific Cluster.  

\bibliographystyle{aa}
\bibliography{refs}

\end{document}